\documentclass[]{aastex62}
\newcommand{\unit}[1]{\ensuremath{\, \mathrm{#1}}}
\newcommand{\Lagr}{\mathcal{L}}
\newcommand{\be}{\begin{equation}}
\newcommand{\ee}{\end{equation}}
\newcommand{\avgc}{\langle\cos2\phi\rangle}
\newcommand{\avgs}{\langle\sin2\phi\rangle}
\received{April 7, 2019}
\revised{June 11, 2019}
\accepted{June 17, 2019}
\published{September, 2019}
\submitjournal{PASP}
\shorttitle{Gamma Ray polarimetry}
\shortauthors{Ilie}

\begin{document}

\title{Gamma Ray polarimetry; a new window for the non-thermal Universe}

\correspondingauthor{Cosmin Ilie}
\email{cilie@colgate.edu}
\author{Cosmin Ilie}
\affiliation{Department of Physics and Astronomy, Colgate University\\
13 Oak Dr., Hamilton, NY 13346, U.S.A.}
\affiliation{Department of Theoretical Physics, National Institute for Physics and Nuclear Engineering\\
 Magurele, P.O.Box M.G. 6, Romania}

\begin{abstract}
Over the past few decades impressive progress has been made in the field of photon polarimetry, especially in the hard X-ray and soft gamma-ray energy regime.  Measurements of the linear degree of polarization for some of the most energetic astrophysical sources, such as Gamma Ray Bursts (GRBs) or Blazars, are now possible, at energies below the pair creation threshold. As such, a new window has been opened for understanding the exact nature of the non-thermal emission mechanisms responsible for some of the most energetic phenomena in the Universe. There are still many open questions and active debates, such as the discrimination between leptonic vs. hadronic models of emission for Blazars or ordered vs random field models for GRBs. Since the competing models predict different levels of linear photon polarization at energies above $~\sim~1\unit{MeV}$, gamma-ray polarimetry in that energy band could provide additional crucial insights. However, no polarimeter for gamma-rays with energies above $~\sim~1\unit{MeV}$ has been flown into space, as the sensitivity is severely limited by a quick degradation of the angular resolution and by multiple Coulomb scatterings in the detector. Over the past few years, a series of proposals and demonstrator instruments that aim to overcome those inherent difficulties have been put forth, and the prospects look promising. The paper is organized as follows: in Sec.~\ref{Sec:History}, I briefly review the history and principles of gamma-ray polarimetry, emphasizing its challenges and successes; Sec.~\ref{S:ProdPol} is dedicated the discussion of gamma-ray polarization and polarimetry, whereas in Sec.~\ref{Sec:Instruments} I discuss the past and current instruments with which measurements of linear polarization for hard X-rays and soft gamma-rays were successfully obtained for astrophysical sources; Sec.~\ref{Sec:Motivations} outlines the scientific questions that could be solved by using gamma-ray polarimetry measurements. I end with a summary and outlook in Sec.~\ref{Sec:Outlook}
\end{abstract}

\keywords{dark matter --- gamma rays: general ---  gamma-ray burst: general --- instrumentation: polarimeters --- polarization  }

\section{Introduction}\label{Sec:History}

\subsection{A Brief history of gamma-ray polarimetry}

Nearly all high-energy emission mechanisms in thermal and relativistic astrophysics (synchrotron, curvature, inverse Compton scattering, magnetic photon splitting) can give rise to linear polarization, with values for the Stokes parameters highly dependent on the source physics and geometry. For instance, for both synchrotron and curvature radiation the degree of polarization is energy independent. However, in the case of inverse Compton scattering the polarization degree depends on energy and the scatter angle. Moreover, based on the orientation of the magnetic field at the origin and the dominant emission mechanism, one can expect different degrees and directions of polarization. It is important to note that for gamma-rays produced by jets of relativistic matter impinging on intra galactic matter via hadronic interactions and subsequent pion decays the expected degree of polarization is zero.  Thus, polarization could play a critical role in disambiguating between currently competing astrophysical models for distribution of magnetic fields, radiation fields, interstellar matter, and emission mechanisms for gamma-rays. In principle, the polarization of electromagnetic waves from any source is a priori arbitrary, i.e. elliptical or circular. Most relevant for astrophysics, is the limit case of linear polarization, where there is a preferred direction for the orientation of the oscillating electric field in the wave. While circular polarization can sometimes be relevant for astrophysical sources,~\footnote{such as for instance the case of GRB 121024A~\citep{GRB121024A}} in general most theoretical models predict low degrees of circular polarization for astrophysical sources. Moreover, the detection techniques for circular vs. linear are somewhat different. Therefore, in this paper I will focus on linear polarization of gamma-rays and hard X-rays. Loosely defined, the threshold between gamma-rays and X-rays is at a few hundred \unit{keV}. Note however, that different authors have different conventions for this purpose. 

Polarimetry complements photometry, spectroscopy, and imaging/mapping, adding information about two physical parameters: the degree and direction of polarization of the incident radiation. It is a ``powerful probe into the gamma-ray emission mechanism and the distribution of magnetic and radiation fields, as well the distribution of matter around wide variety of astro sources''~\citep[PoGOLite proposal:][]{Kamae:2008PoGOLite}. For instance, it could shed light on the emission mechanism in pulsars, X-ray binary systems, jets in AGN and microquasars, just to name a few. The first astronomical observations of polarized light were achieved in the middle of the 19th century. For example, seminal work on the linear polarization of the sunlight reflected by the moon~\citep{Secchi:1860} and the linear polarization of the light from solar corona~\citep{Edlund:1860} was published during that period. Some of the major milestones in science that were obtained using  polarimetry are: the discovery of synchrotron radiation from the Crab nebula~\citep{Oort:1956}, the study of the surface composition of solar system objects \citep{Bowell:1974}, the measurement of the X-ray linear polarization of the Crab nebula~\citep{Weisskopf:1978} - which is still one of the best measurements of linear polarization for astrophysical sources, mapping of solar and stellar magnetic fields~\citep[see][ch. 4]{Schrijver:2000}, the detection of polarization in the Cosmic Microwave Background (CMB) radiation~\citep{Kovac:2002}, the direction and curvature of large scale, e.g. $\unit{kpc}$, galactic magnetic fields~\citep[e.g][]{Heiles:1996,Kulsrud:2007}, etc. For a review on the status of photon polarimetry see \citet{Trippe:2014}.

 \subsection{Challenges and successes of photon polarimetry}\label{SS:Challenges}

Two of the most important parameters of any astronomical telescope are its angular resolution and sensitivity. For photon detection, the highest sensitivities  are achieved at low energies ($<100 \unit{keV}$) and at high energies ($>100 \unit{MeV}$); however polarimetry for gamma-rays beyond the pair production threshold is still very challenging for reasons I will explain in detail in Sec.~\ref{S:ProdPol}. In addition, the polarization signal is usually just a small part of the total radiation emitted by any given astrophysical source, and therefore, photon polarization measurements have been possible only for the brightest sources. Despite the inherent challenges, in recent years, some instruments have achieved important milestones in hard X-ray and soft gamma ray polarimetry, as discussed below, and in more detail in Sec.~\ref{SSec:Past}. Unfortunately, no polarimeter sensitive to photons of energies above $\sim 1\unit{MeV}$ has been flown in space yet. However, in the near future this situation is bound to change, as numerous proposals have been put forth. One approach is based on advances in Silicon detector technology. Here are two such examples: the e-ASTROGAM mission~\citep{DeAngelis:2016} or the High Energy Photon Polarimeter for Astrophysics proposed by ~\citet{Eingorn:2018JATIS}. Other possible approaches are gas time-projection chamber (TPC) polarimeters, such as the HARPO project proposed by ~\citet{Bernard:2012} or emulsion based detectors such as the GRAINE project~\citep{GRAINE:2015}. For more details, please see Sec.~\ref{SSec:Proposals}

\begin{figure*}
 \begin{center}$
 \begin{array}{c}
 \includegraphics[scale=0.5]{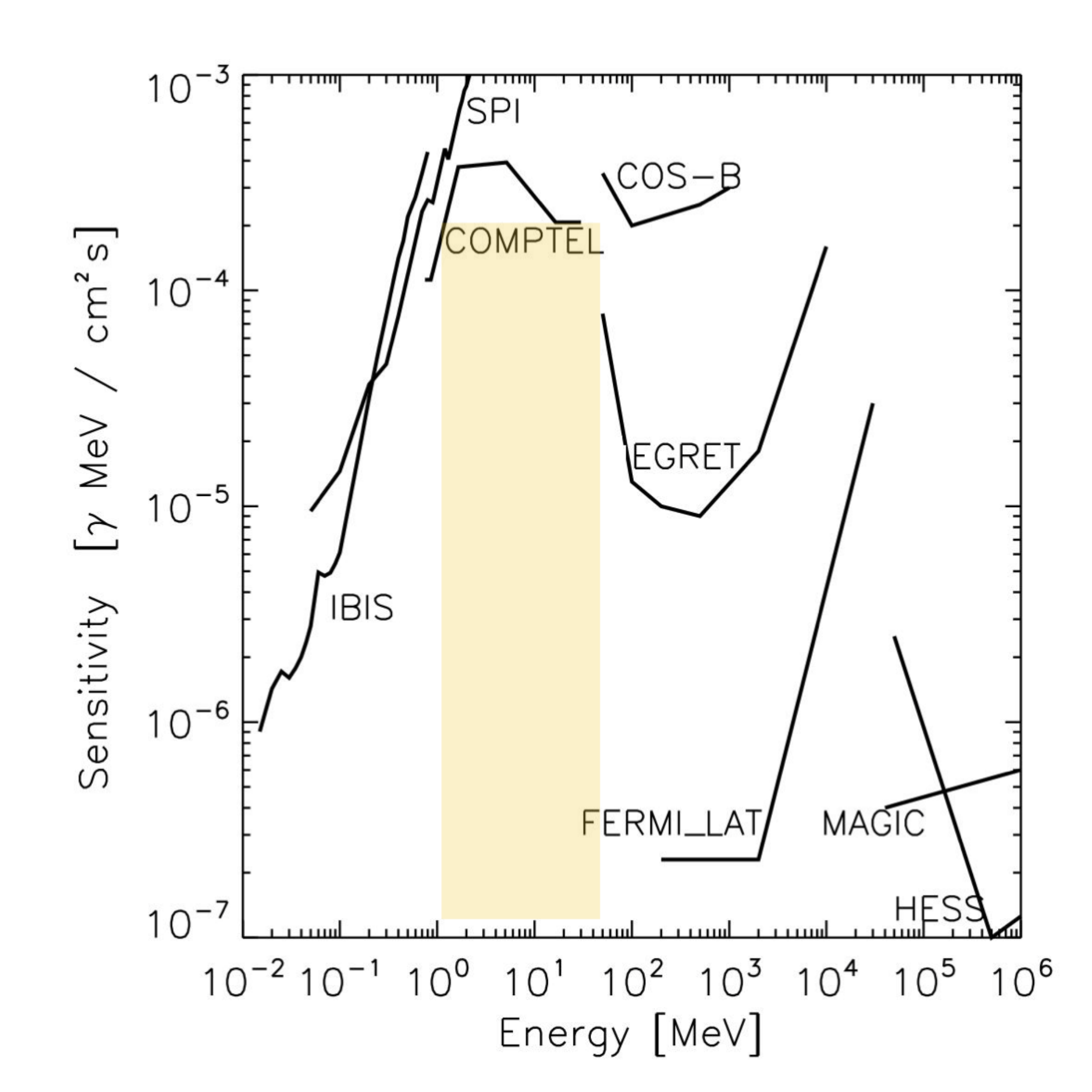}
 \end{array}$
 \end{center}
 \caption{Point source sensitivity in the gamma-ray domain achieved by various previous or currently operational instruments. Note the shaded region at energies of a few $\unit{MeV}$ up to roughly $100\unit{MeV}$. Both Compton scattering and pair production are relevant at this energy for most detectors, and therefore background rejection becomes extremely challenging due to the quick degradation of angular resolution. This is the main reason behind the so called ``$\unit{MeV}$ sensitivity gap,'' the shaded region of this plot.}
 \label{Fig:SensGap}
 \end{figure*}

Even regarding photon detection alone, without any information regarding polarization, there exists  a wide instrumental point source sensitivity gap in the energies of above the pair creation threshold ($\sim1\unit{MeV}$) to around $50 \unit{MeV}$, as one can see in Fig.~\ref{Fig:SensGap}.  The main problem in this region is the challenging nature of rejecting background due to poor angular resolution of single photon measurements. Numerous astrophysical sources, for which the exact nature of their emission mechanisms remains poorly understood, have peak emissivity in this important transition region, between thermal and non thermal processes. To fill this sensitivity gap, scientists are testing advanced Compton telescopes  with different detection concepts, such as  silicon strip detectors, position sensitive germanium, CdTe, liquid xenon gas, and high-pressure gas detectors instead of scintillators \citep{Schonfelder:2004}. In the remainder of this  section, I will briefly discuss some of the major recent milestones in measurements of polarization of hard X-rays and soft gamma-rays of astrophysical origin.

 %%%%%%%%%%% Figure Sens Gap%%%%%%%%%%%%%%%

The Gamma-Ray Polarization Experiment (GRAPE) is a balloon borne polarimeter developed in order to measure polarization for astrophysical soft gamma-ray sources at low energies ($50-500 \unit{keV}$). This instrument could be used for polarization measurements of gamma-rays from outbursts of energetic astrophysical objects such as GRBs and solar flares \citep{Bloser:2005Grape}. In 2011, GRAPE was tested and performed observations of the Crab Nebula and two M-Class solar flares during 26 hours at float altitude \citep{Bloser:2013Grape}. The Polarized Gamma-ray Observer ~\citep[PoGoLite:][]{Kamae:2008PoGOLite} is another baloon-borne instrument designed to provide higher sensitivity for the point sources in the $25-240 \unit{keV}$, but not for GRBs. For a period of two weeks during July 2013, PoGOLite was used to observe the Crab nebula while in a near-circumpolar pathfinder fight from Esrange, Sweden to Norilsk, Russia. It detected the characteristic pulsation in the light curve due to the Crab pulsar \citep[see][]{Kawano:2015PoGOLite}. Moreover, the PoGOLite pathfinder mission was able to successfully measure polarization of hard X-ray emission from the crab, as reported by ~\citet{Chauvin:2015}. This is a very significant result, since it was the first time that polarization emission was measured in the hard X-ray band for astrophysical sources! However, PoGOLite, and its successor, PoGO+, are only sensitive to linear polarization below $160 \unit{keV}$. Therefore, a very large swath of the high energy photon polarimetry remains unexplored for astrophysical sources.

For medium energies, NASA's Imaging Compton Telescope (COMPTEL), that was on board the Compton Gamma Ray Observatory (CGRO) during 1991-2000 and ESA's International Gamma-Ray Astrophysics Laboratory (INTEGRAL) with its IBIS imager and SPI spectrometer, although not designed as polarimeters, are capable of polarimetry. This is achieved by cleverly exploiting the dependence of the Compton scatter angle on the polarization of the incoming photon. For the measurements of the polarization from the Crab nebula using IBIS see~\citet{Forot:2008}. For a more recent analysis that is based upon data from nearly ten years of operation of INTEGRAL-IBIS see~\citet{Moran:2013Crab}. Both studies agree in finding that off-pulse emission from Crab is highly polarized, with a polarization angle aligned with the rotation axis of the pulsar. Another target for which polarization was successfully detected with INTEGRAL-IBIS is the black hole X-ray binary system Cygnus X-1. \citet{Laurent:2011} found that in the $200~\unit{keV}-2~\unit{MeV}$ spectral band the gamma-rays from Cygnus X-1 is strongly polarized ($\Pi\sim 67\pm30\%$), whereas emission in the $250-400~\unit{keV}$ band is very weakly polarized, indicating different emission mechanisms. Those results were confirmed by ~\citet{Jourdain:2012}, who developed the required tools to study the polarization in the INTEGRAL SPI data and applied them to Cyg X-1. Compton scattering on thermal electrons is consistent with the low polarization signal, whereas the high polarization at higher energies could be explained either by synchrotron emission or by Inverse Compton scattering from the jet that was already observed in the radio band \citep{Fender:2004}.

INTEGRAL has the capability to detect the signature of polarized emission from bright  gamma-ray sources, such as GRBs. So far, such measurements have proven to be very challenging due to low $S/N$ ratios. However, there have been several encouraging results in the past decade, such as the variable polarization measured in the prompt gamma-ray emission from the bright GRB 041219A with IBIS by ~\citet{Gotz:2009GRB} or with SPI by~\citet{McGlynn:2007}, just to name a few. For an overview of recent polarimetric observations obtained with INTEGRAL, see~\citet{Laurent:2016}. In Sec. ~\ref{Sec:Instruments}, I will discuss in more detail the use of IBIS or SPI as a polarimeters, its limitations, and a non-exhaustive list of GRBs for which polarization was detected using those two instruments on board INTEGRAL. Polarization measurements  from COS-B, CGRO/EGRET, FERMI-LAT (formerly known as GLAST), and AGILE have been unsuccessful at high energies ($E_{\gamma}>30 \unit{MeV}$). Multiple Coulomb scattering limits effective polarization measurements from  pair production telescopes, as background rejection and event selection become increasingly difficult. Furthermore, present (FERMI/LAT) and previous(COS-B, EGRET) $e^{+} e^{-}$ pair telescopes have no significant sensitivity to the polarization of the incoming photons above $\sim\unit{MeV}$. It is therefore essential to design a new pair creation polarimeter dedicated to obtaining polarization measurements for medium and high energy gamma-rays.

Low energy photon polarimetry is a very active field of research, and great progress has been made in the past 40 years. One of the most exciting results was the detection  of B mode polarization due to lensing in the CMB with SPT~\citep{Hanson:2013}. The detection of primary B and E modes would lead to a better understanding of the physics of the early universe and could place severe constraints on models of cosmic inflation. Pulsars, galactic magnetic fields, the geometries of magnetic fields and particle densities around active galactic nuclei (AGN)~\citep{Saikia:1988}, AGN jets~\citep{Macquart:2006,Taylor:2006}, and cosmic shock waves~\citep{Clarke:2006} represent just a few of the numerous directions of research where polarization in the radio band has played a significant role. One very impressive result is the creation of all sky polarization  maps using the 21 $\unit{cm}$ line~\citep{Testori:2008}. Plans for the near future include the use of the Square Kilometer Array (SKA) instrument for wide-field radio polarimetry, as proposed by ~\citet{Gaensler:2015}. This technique could, in principle, help address some of the mosts important  questions in astrophysics and cosmology, including the relationship between supermassive black holes and their environment or how galaxies have evolved over time. In a landmark result, ALMA has recently been used to detect and measure the temporal evolution of polarized radio/milimeter emission from the gamma-ray burst GRB 190114C~\citep{Laskar:2019}, which offers new insights into understanding the GRB phenomena. In the the optical, UV and X-Ray wavelength bands polarimetry was used for astrophysical sources for decades. Some examples where polarimetry played an important role in discoveries at those energies include: detection of exoplanets~\citep{Berdyugina:2007}; chemical composition of planetary atmospheres~\citep[e.g.][]{Stam:2004}; the study of interstellar matter~\citep[e.g.][]{Wilking:1980}; quasar jets~\citep{Cara:2013}; and solar flares~\citep{Boggs:2006,Suarez-Garcia:2006}.

At higher energies, X-ray or Gamma-ray polarimetry have inherent challenges that stem from the low flux of typical astrophysical sources at such high frequencies. This usually implies the necessity to be able to observe individual photons, placing very tight technological constraints in the medium to high gamma-ray energy regime. Despite of those limitations, there are numerous polarimetry studies in X-ray and soft gamma-ray astrophysics, with various proposals put forth regarding medium and high energy gamma-ray polarimeters. This will be discussed at length in Sec.~\ref{Sec:Instruments}. In Sec.~\ref{Sec:Motivations}, I will present the main motivations for a polarimeter in the $\unit{MeV}-\unit{GeV}$ regime, the various emission mechanisms of gamma-rays at those energies, and how polarimetry could be used to disambiguate among them. Sec.~\ref{Sec:Outlook} presents a summary and outlook on the perspectives of gamma-ray polarimetry in the $\unit{MeV}-\unit{GeV}$ band.

\section{Gamma-ray polarization and polarimetry}\label{S:ProdPol}

\subsection{Production of Gamma-Rays and polarization mechanisms}\label{SSec:ProdMech}

Thermal production of gamma-rays is negligible in most astrophysical sites, as the temperature required for a black body thermal radiator to have a significant emission in the $\sim\unit{MeV}$ energy is of the order of $10^{10}\unit{K}$! For example, our Sun has a core temperature three orders of magnitude below that. Therefore, non-thermal processes are important in the production of galactic and extra-galactic gamma-rays.

\begin{figure*}
\begin{center}$
\begin{array}{c}
\includegraphics[scale=0.4]{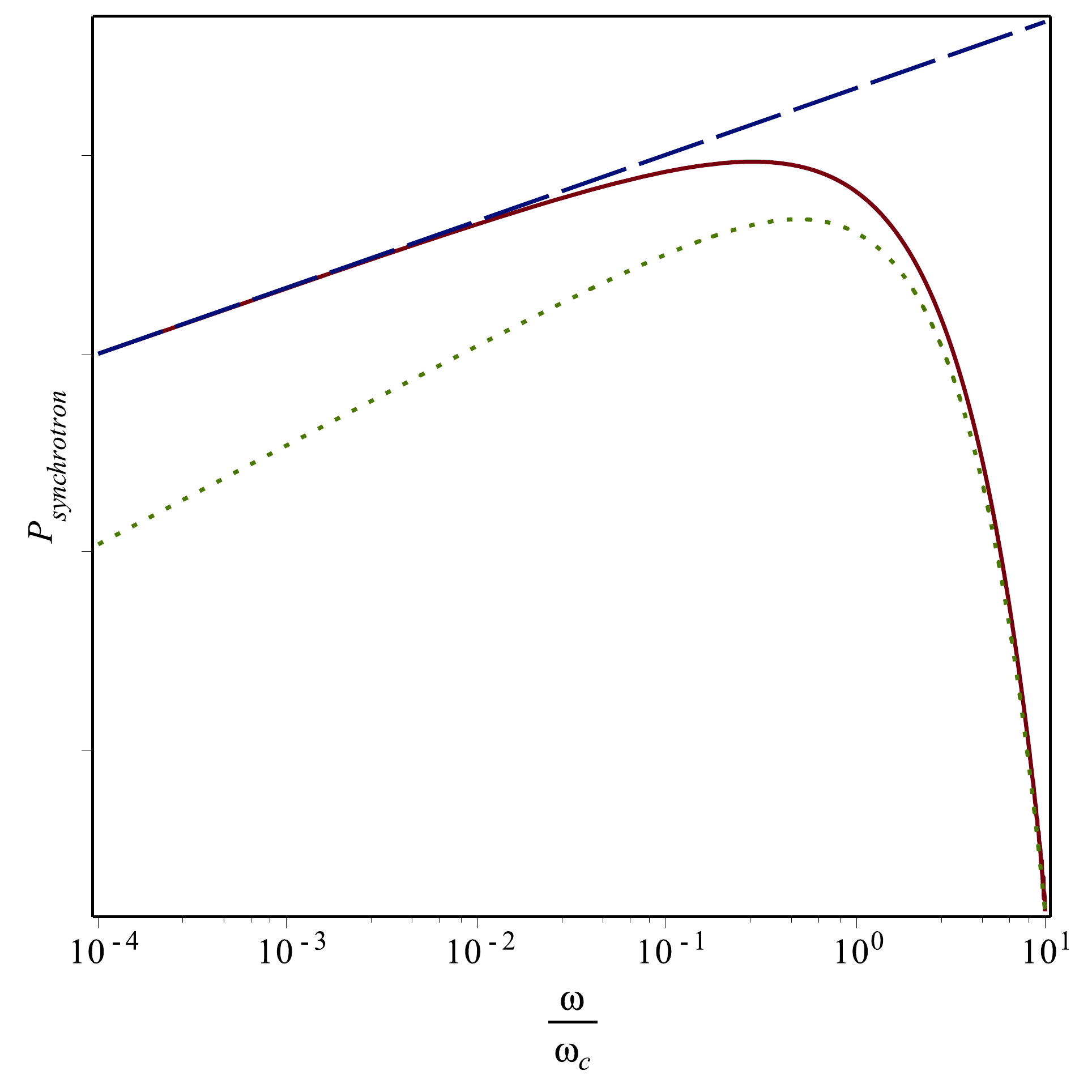}
\end{array}$
\end{center}
\caption{Log log plot of synchrotron power distribution as a function of frequency. The solid (red) curve represents the theoretical spectrum. At low frequencies this can be very well approximated by a power law, i.e. $P\sim \omega^{1/3}$, as depicted by the dashed (blue) line. At the high frequency end, the spectrum exhibits a sharp exponential cutoff in the power emitted, as can be seen from the dotted (green) curve: $P\sim\omega^{1/2}\exp(-\frac{\omega}{\omega_c})$}
\label{Fig:SRPower}
\end{figure*}

{\bf{Synchrotron Radiation:}} Of those processes, one of the most ubiquitous is the acceleration of charged particles in strong magnetic fields, leading to synchrotron (when the motion is primarily circular, around the field lines) or curvature (if the motion is primarily along the field lines) radiation, both of which are highly polarized~\citep{Westfold:1959}. Typical astrophysical sites are Supernovae Remnants (SNRs), pulsars, and active galactic nuclei (AGNs). We note here that for young recycled pulsars the transition between synchrotron radiation and curvature radiation is marked by a flip in polarization below $100~\unit{MeV}$, as shown by~\citet{Harding:2017}.  The interested reader can find more details regarding the mechanisms and various sources of synchrotron radiation in~\citet{Wille:1991}. Synchrotron radiation was first invoked as a possible explanation for polarized radiation by Russian physicists shortly after the discovery of polarized radio emission from SNRs and radio galaxies. For example,~\citet{Shklovskii:1957} demonstrates that supernova remnants are synchrotron emitters by explaining the optical polarization of the Crab Nebula using this assumption. Physically, the same processes that happen in astrophysical sites for ultra-relativistic particles that traverse regions with significant magnetic fields are responsible for the radiation observed from relativistic particles in manmade accelerators that use strong magnetic fields to guide electrically charged particles in closed paths. The geometry of the radiation depends on the energy of the accelerated particle, changing from dipolar, for non-relativistic electrons, to radiation being beamed into a cone of angle $\theta\simeq m_ec^2/E=1/\gamma$, for relativistic electrons. As one can see from Fig.~\ref{Fig:SRPower}, the power emitted by synchrotron radiation exhibits a sharp cutoff at frequencies higher than a critical value:

\be\label{ESRCritFreq}
\omega_c=(3/2)(eB/mc)\gamma^2\sin\phi,
\ee
with $\phi$ being the pitch angle between the direction of the magnetic field and that of the electron.

%%%%%%%%%%% Figure %%%%%%%%%%%%%%%

The degree of polarization and spectral distribution of synchrotron radiation was first calculated by \citet{Westfold:1959}. He has shown that the degree of polarization in the emission increases with frequency, from two-thirds at radio frequencies to unity at high frequencies, corresponding to gamma-rays. Depolarization effects, due to the non-uniformity of the magnetic fields, are responsible for a lesser degree of polarization in the denser central regions as opposed to the edges. This leads to a measured value for the polarization degree being less than unity, even for the highest energy photons emitted by electrons gyrating ultra-relativistically in regions where the magnetic fields varies in both magnitude and direction. The observed radiation is the convolution of the single-particle spectrum with the distribution of the relativistic electrons as a function of energy. Over a wide rage of energies one can approximate this distribution as a power law: $n(E)dE\sim E^{-k}dE$, where $k$ is the spectral index. The radiation is highly elliptically polarized, i.e. very nearly linear. One defines the degree of polarization in terms of the radiation powers $P_{\parallel}$ and $P_{\perp}$, i.e. the power in the radiation with the electric field vector parallel, respectively perpendicular, to the projection of the magnetic onto the plane of the sky:

\be
\Pi(\omega)=\frac{P_{\perp}(\omega)-P_{\parallel}(\omega)}{P_{\perp}(\omega)+P_{\parallel}(\omega)}
\ee

For the case of power law synchrotron radiation, considered above, the degree of polarization integrated over all frequencies and all electron energies is given by~\citep{Ginzburg:1969,Pacholczyk:1970,Rybicki:1985}:

\be\label{ESRDegPol}
\Pi=\frac{3k+3}{3k+7}.
\ee

For a typical value of $k\sim 2.4$, this translates to a degree of polarization of $\Pi\sim 0.72$, which is quite high. Moreover, the linear polarization shows the projected direction of the magnetic fields, thus making synchrotron emission polarimetry a very powerful tool for studying astrophysical magnetic fields. In order to have a significant gamma-ray component of the synchrotron radiation either the magnetic fields need to be very strong or the electrons ultra-relativistic, due to the exponential cutoff of the emitted power at frequencies beyond the critical frequency, given by Eq.~\ref{ESRCritFreq}.

 {\bf{Bremsstrahlung Radiation:}} Another important mechanism for non-thermal production of gamma-rays is bremsstrahlung of electrons being slowed down by electromagnetic interactions with nuclei. For a single electron in the presence of a positive charge, the radiation pattern transitions between a constant intensity at low frequencies, and a relatively sharp cutoff, imposed by conservation of energy. Essentially, an electron cannot radiate more energy than its kinetic energy. \citet{Gluckstern:1953} calculated the polarization of the bremsstrahlung radiation and found that it consists of a mixture of unpolarized and linearly polarized components. In a completely ionized plasma in thermal equilibrium, near collisions between electrons and ions lead to what is called thermal (or relativistic) bremsstrahlung, depending on the velocity of the electrons. Detailed calculations, integrating the single-electron spectra over all possible electron velocities lead to a thermal bremsstrahlung spectrum. This has three distinct regions: a) $I(\omega)\propto\omega^2$, at low frequencies, where the process of ``self-absorption'' becomes important; b) $I(\omega)\propto const$, at intermediate frequencies; and c)  $I(\omega)\propto e^{-h\omega/kT}$, the thermal cutoff at high frequencies. All of those features are due to the Maxwell-Boltzmann distribution of electron velocities.

In astrophysics, thermal bremsstrahlung is an important source of X-rays and soft gamma-rays from SNRs, solar flares, and Inter Stellar Medium (ISM). In general, the detection of bremsstrahlung radiation indicates a source with an important ionized gas or plasma component, such as stellar atmospheres, and the central regions of AGNs (or other objects accreting matter). Regions of ionized hydrogen (HII) at temperatures of $T\sim 10^4$ K are known sources of bremsstrahlung radiation at radio frequencies, whereas hot intergalactic gas in cluster of galaxies is a source of diffuse X-ray emission via thermal bremsstrahlung. Because of the random motion directions of the electrons generating thermal bremsstrahlung, this radiation is usually assumed to be unpolarized. However,~\citet{Komarov:2016} considered the intriguing possibility that electron pressure anisotropy could lead to a polarization of the thermal bremsstrahlung radiation.

{\bf{Inverse Compton Scattering}} (ICS) happens when low-energy photons are scattered up to X-Ray and gamma-ray energies by relativistic particles. This process is efficient only in sites where there is a high photon density. Moreover, as with the previous non-thermal processes listed, it relies on the presence of an acceleration mechanism. Examples of sites where ICS is present include: pulsars, SNR, X-Ray binaries, ISM, AGN, and IGM. The theory for evaluating polarization of the Inverse Compton scattered radiation was first developed by~\citet{Bonometto:1970} and applied to the case of synchrotron self-Compton (SSC) radiation by~\citet{Bonometto:1973}, in the low energy Thomson regime. In the relativistic regime~\citet{Krawczynski:2012} provides a general Monte-Carlo approach for the evaluation of polarization signatures. Moreover,~\citet{Krawczynski:2012} verified numerically that the analytical expressions of~\citet{Bonometto:1973} provide excellent estimates for the polarization in the Thomson regime. However, it is worth noting that due to the large Doppler boosting factors typical for blazar sources ($\delta\sim 10$), the polarization of observed SSC radiation can be safely described using the analytical estimates of ~\citet{Bonometto:1973}, even for photons  with energies as high as $0.5\unit{GeV}$.

 {\bf{Nuclear transition, annihilation, and decay lines:}} At $\unit{MeV}$ energies, nuclear production of gamma-rays is important, which manifests itself as distinctive transition lines in the spectra. Particle physics allows for production of gamma-rays by two main channels: $e^+e^-$ annihilation ($511 \unit{keV}$ line  for $e^+e^-\to 2\gamma$, or continuum from $0-511\unit{keV}$ in the case of $e^+e^-\to 3\gamma$) and pion bump ($E\gtrsim 68\unit{MeV}$). Examples of celestial sources where particle physics reaction production of gamma-rays are important include: solar flares, Novae, SNRs, Cosmic Rays interacting with ISM, and the galactic center. It has been proposed in the literature that searches for gamma-ray line signals from the center of the galaxy could be used as a smoking gun signature of Dark Matter annihilations. This is because there are no known astrophysical sources that could generate a sharp, monochromatic signal at energies $\gtrsim 100\unit{MeV}$. For a review of gamma ray signals from dark matter in terms of concepts, status, and prospects, see~\citet{Bringmann:2012b}. There were such hints of detection of a monochromatic gamma-ray line at $\sim 130\unit{GeV}$ found in Fermi-LAT data \citep[see, e.g.][]{Bringmann:2012,Weniger:2012}. The implications of this result in terms of DM annihilation were pointed out by \citet{Buckley:2012}, among others. Unfortunately, the statistical significance of the $\sim 130 \unit{GeV}$ gamma-ray line found in the Fermi-LAT has faded away in recent years. However, another excess at a few $\unit{GeV}$ was identified by~\citet{Goodenough:2009} and confirmed most recently by ~\citet{Ackermann:2017}. The exciting possibility that this signal is due to annihilations of DM particles in the innermost regions of our galaxy has not been excluded; yet there exist other plausible explanations of astrophysical nature, such a millisecond pulsars. Polarization studies could prove to be critical in disambiguating between particle physics signals (DM) and competing astrophysical sources that can mimic the same spectrum, as explained in some detail in Sec.~\ref{Sec:DM}.

In general, the proposed models for many sources of gamma-rays predict very different polarization signatures. Various degrees and directions of polarization are expected, depending on the orientation  of the magnetic field at the source and on the primary emission mechanism. Consequently, gamma-ray polarimetry  can be used to probe the nature and geometry of many objects such as pulsars, binary systems, solar flares, gamma-ray bursts (GRB), and active galactic nuclei.  In Sec.~\ref{Sec:Motivations}, I will present more detailed reasons as to why polarization measurements are essential for high energy ($\unit{MeV}-\unit{GeV}$) astrophysics and what we hope to learn from them.

\subsection{Measuring principles for gamma ray polarimetry}\label{SS:Measuring}

\begin{figure*}
\begin{center}$
\begin{array}{c}
\includegraphics[scale=0.7]{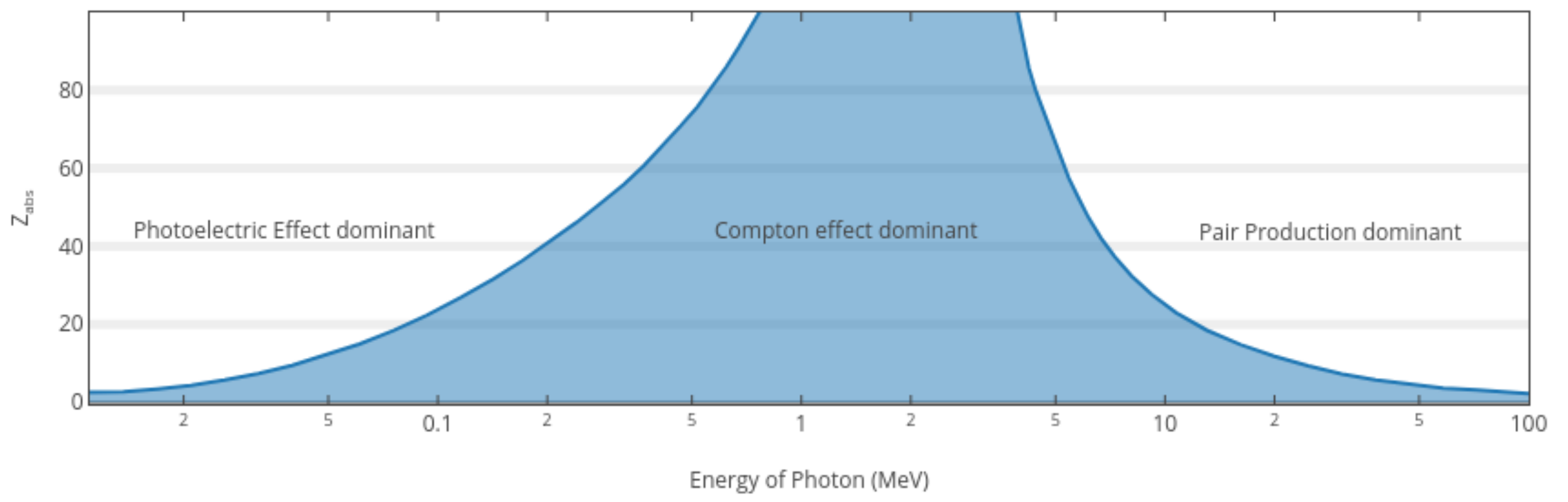}
\end{array}$
\end{center}
\caption{Regions where one of the three $\gamma$-interactions dominates over the other two.It can be seen that at low energies ($\lesssim 1 \unit{keV}$) the photo-electric effect is dominant, at mid energies ($\sim 1 \unit{MeV}$) the Compton effect becomes the most important scattering mechanism, whereas at high energies ($\gtrsim$ few $\unit{MeV}$) pair production becomes significant. Any Compton based polarimeter at high energies suffers from a degradation in the resolution due to multiple Coulomb scatterings in the detector.}
\label{Fig:Regions}
\end{figure*}

There are several distinct ways to measure polarization of photons, and the techniques applied depend largely on the degree of polarization and the energy of the photons (See Fig.~\ref{Fig:Regions}).

%%%%%%%%%%% Figure Regions abs Z and Energy%%%%%%%%%%%%%%%

\begin{figure*}
\begin{center}$
\begin{array}{c}
\includegraphics[scale=0.65]{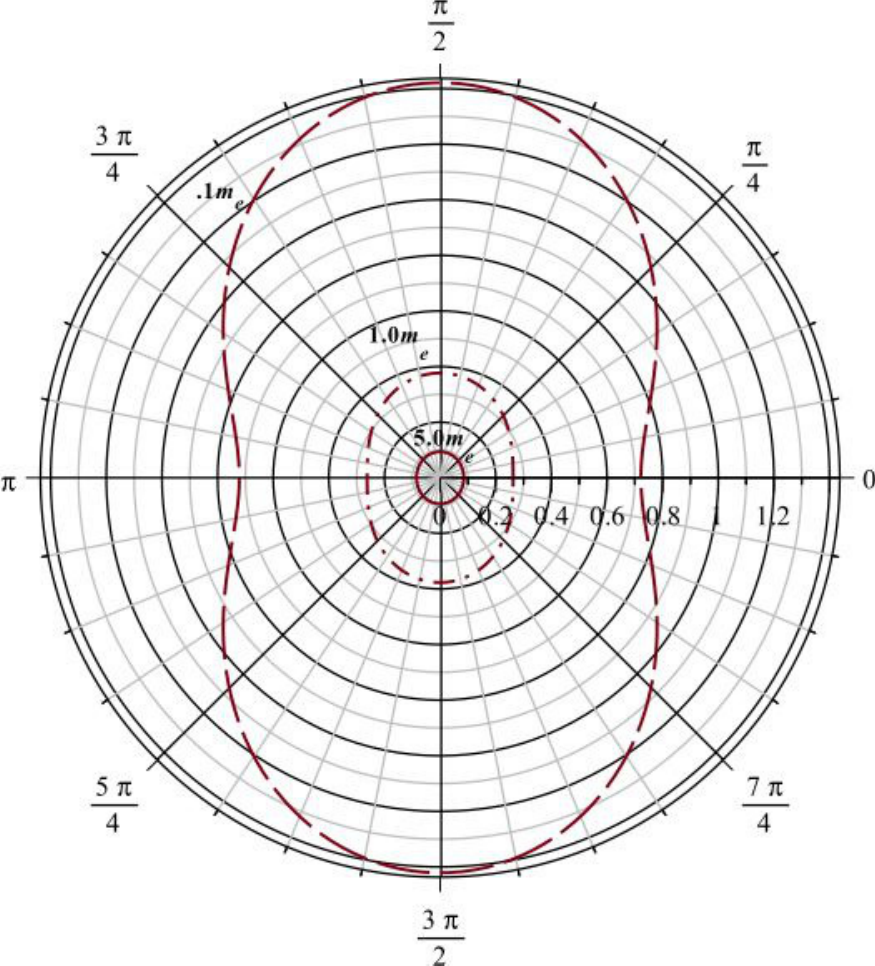}
\end{array}$
\end{center}
\caption{The Klein-Nishina scattering cross section, for a fixed scattering angle, chosen here to be $\theta=\pi/4$ and assuming the incident photon is polarized along the x axis. The lines are labeled with the energy of the incident photon, in multiples of the electron rest mass ($511 \unit{keV}$). Note how the azimuthal asymmetry decreases with increasing energy of the incident photon.}
\label{Fig:KN}
\end{figure*}

At low energies, of the order of $\unit{keV}$, Thompson scattering, the photoelectric effect, and Bragg-reflection are the dominant interactions between gamma-rays and matter. At about $0.1 \unit{MeV}$, Compton scattering starts to dominate. It is worth mentioning that all successful polarization measurements to date for medium and high energy gamma-rays of astrophysical origin are based a measurement of the direction of the Compton scattered photons. Connection to polarization can be made based on the fact that linearly polarized photons Compton scatter preferentially orthogonal to the direction of the polarization vector, i.e. at $90\degr$ angles to the direction of the electric field. This can be seen from the analytic expression of Klein-Nishina differential cross section~\citep{KN:1929}, which for polarized incident photons reads:

\be\label{Eq:KN}
\frac{d\sigma_{KN,P}}{d\Omega}=\frac{3}{16\pi}\sigma_T\epsilon(\theta)^2\left[\epsilon(\theta)+ \epsilon(\theta)^{-1}-\sin^2\theta\cos^2(\eta-\phi)\right].
\ee
The quantity labeled $\sigma_T$ in Eq.~\ref{Eq:KN} represents the Thomson cross section, $\epsilon$ is the ratio between the scattered and the incident photon energies (see Eq.~\ref{Eq:RatioE}), $\eta$ is the azimuthal scattering angle, and $\phi$ is the polarization angle of the incident photons. In Fig.~\ref{Fig:KN}, I plot the differential cross section for a fixed scattering angle, for which a value of $\theta=\pi/4$ is chosen. It is clear that at low energies the photons scatter preferentially orthogonal to the direction of polarization of incident photons. At the same time, the azimuthal asymmetry almost vanishes at energies of a few $\unit{MeV}$, which is one of the most severe limitations of the use Compton gamma-ray polarimeters at energies above the pair creation threshold.

% %%%%%%%%%%% Figure KN   F1%%%%%%%%%%%%%%%
\begin{figure*}
 \begin{center}$
 \begin{array}{c}
 \includegraphics[scale=0.35]{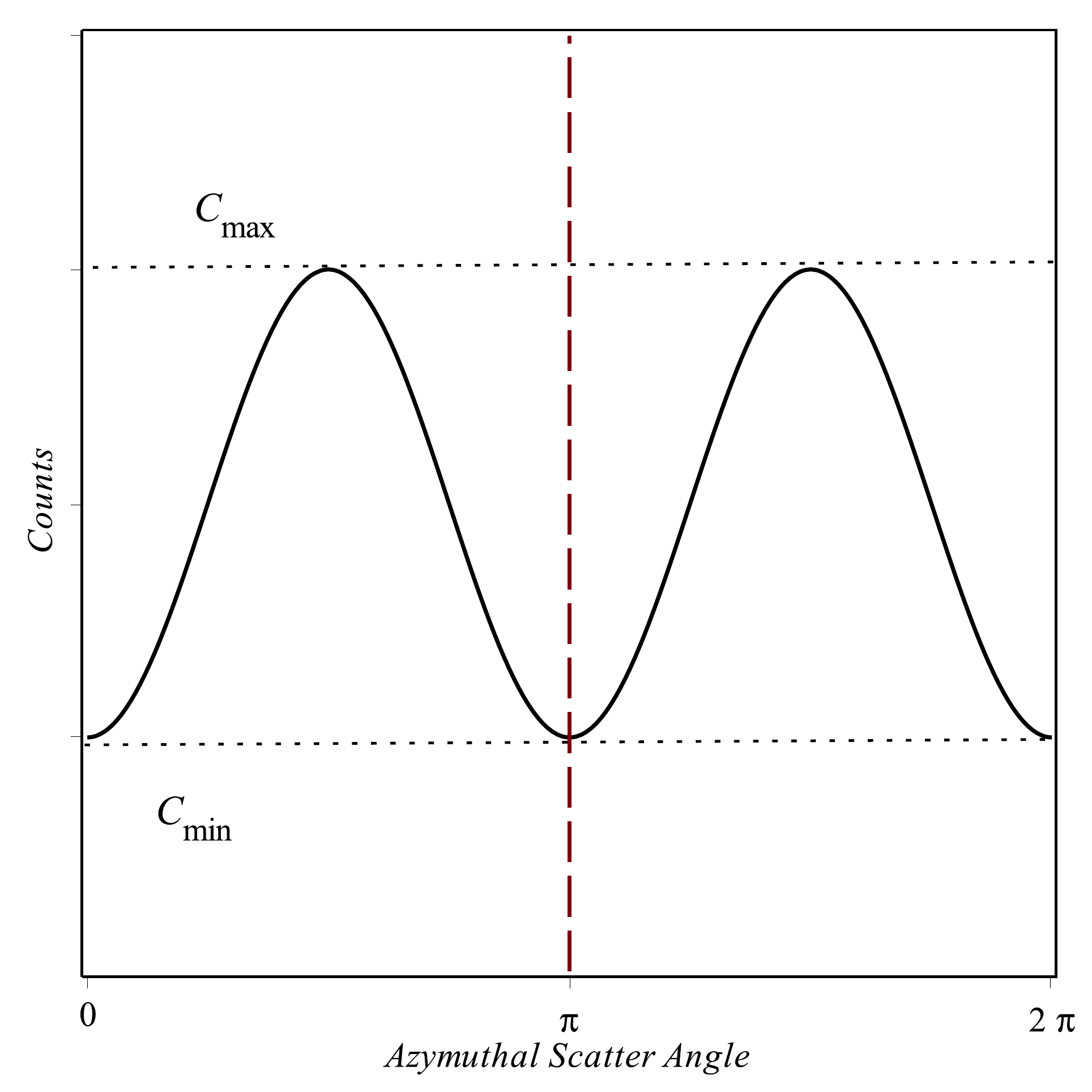}
 \end{array}$
 \end{center}
 \caption{The theoretical azimuthal modulation pattern for Compton scattering polarized radiation, as described by the fit function in Eq.~\ref{Eq:ComptCounts}. The dashed vertical line through the minimum of the distribution corresponds to the polarization angle, in this case $\pi$.}
 \label{Fig:AzCompton}
 \end{figure*}

%%%%%%%%%%%%%%%%%%%%%%%%%%%%%%%%%%%%%%%%%
At energies below a few $\unit{MeV}$, the azimuthal modulation of the distribution of scattered photons can be used to extract information regarding linear polarization of the incident gamma-ray. The modulation pattern can be fit with the following function:

\be\label{Eq:ComptCounts}
C(\eta)=A\cos(2(\eta-\phi+\frac{\pi}{2}))+B,
\ee
where $C(\eta)$ is the number of counts as a function of the azimuthal angle, $\eta$, $\phi$ is the polarization angle of the incident photons, and $A$ and $B$ are fit parameters. This fit function is plotted in Fig.~\ref{Fig:AzCompton}. Note how the angle (plane) of polarization is found from the minimum of the distribution. Moreover, from the amplitude of the modulation, one can find the level of polarization, i.e. the polarization  modulation factor $\mu_p$:
\be\label{Eq:PolModFact}
\mu_p=\frac{C_{max}-C_{min}}{C_{max}+C_{min}}=\frac{A}{B},
\ee
with $A$ and $B$ being the fit parameters introduced in Eq.~\ref{Eq:ComptCounts}.

 %%%%%%%%%%% Figure Compton Polarim F2%%%%%%%%%%%%%%%

In order to extract the polarization information from a particular experiment, one needs to know the response of the detector to a beam of $100\%$ polarized photons and its associated modulation factor $\mu_{100}$. This can be found either by calibrating it experimentally or by using Monte Carlo simulations for the particular detector design. In the literature, this quantity is also known as the Q polarimetric modulation factor, or simply the Q factor. 

\begin{figure*}
 \begin{center}$
 \begin{array}{c}
 \includegraphics[scale=0.45]{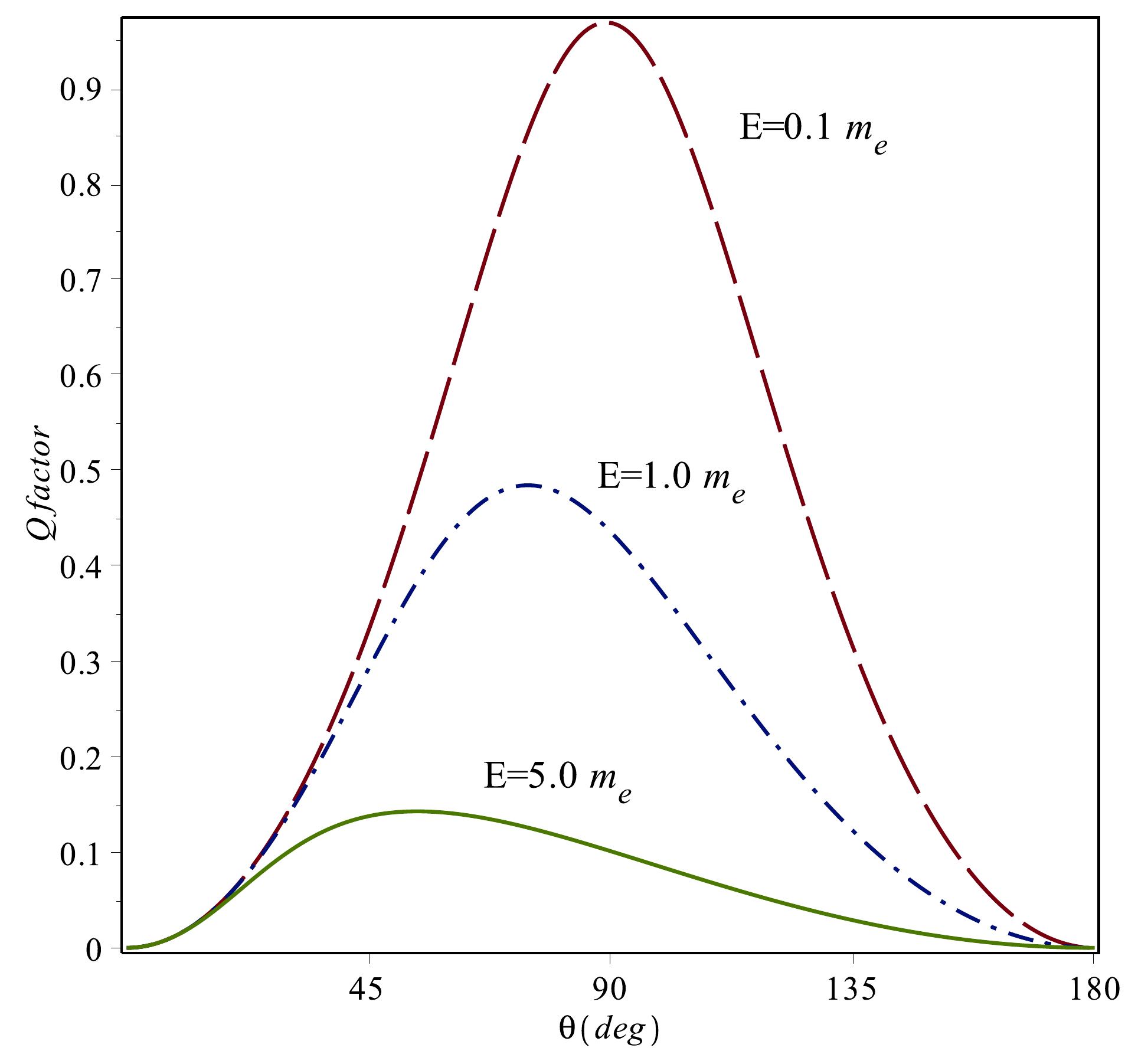}
 \end{array}$
 \end{center}
 \caption{Q polarimetric modulation factor as a function of scattering angle for the case of an incident photon $100\%$ linearly polarized. The three lines correspond to different incident energies, labeled on top of each curve. They match the energies considered in Fig.~\ref{Fig:KN}}
 \label{Fig:QFact}
 \end{figure*}

For Compton scattering of $100\%$ polarized light, one can compute theoretically what is the upper bound (i.e. for an ideal detector) for $\mu_{100}$. Using the Klein-Nishina formula for the differential cross section in Eq.~\ref{Eq:KN} we get:

\be\label{Eq:QFact}
Q\equiv\frac{N_{\bot}-N_{\parallel}}{N_{\bot}+N_{\parallel}}\stackrel{K-N}{=}\frac{\sin^2\theta}{\epsilon(\theta)+\epsilon(\theta)^{-1}-\sin ^2\theta},
\ee
where $N_{\bot}$ and $N_{\parallel}$ represent the number of counts in a direction orthogonal or parallel respectively, with respect to the direction of polarization of the incident photon. In Fig.~\ref{Fig:QFact}, I plot the dependence of the Q factor with the scattering angle for various incident photon energies, measured as multiples or submultiples of the electron rest mass, $m_e$. It is immediately clear that the modulation fraction decreases with energy. Hence, it renders a Compton telescope insensitive to polarization for energies above a few $\unit{MeV}$, as we have already seen from analyzing the azimuthal modulation of the Klein-Nishina scattering cross section, plotted in Fig.~\ref{Fig:KN}. One can show that at high energies $Q\sim 2m/E$. This can be derived from expression for the Q polarimetric factor from Eqn.~\ref{Eq:QFact}, combined with the expression for ratio between the scattered and the incident photon energies:
\be\label{Eq:RatioE}
\epsilon(\theta)=\frac{1}{1+\frac{E}{m_e}(1-\cos\theta)}.
\ee
At high energies photons tend to scatter forward, as demonstrated by the transition of the peak in Q factor towards lower scattering angles in Fig.~\ref{Fig:QFact} as the incident photon energy is increased. Therefore, to a good approximation one can keep only leading order terms in $\theta$, which reproduces the result $Q\sim 2m/E$.

 %%%%%%%%%%% Figure QFactor%%%%%%%%%%%%%%%

The polarization $\Pi$ is computed from the Q factor (i.e.$\mu_{100}$) in the following way:
 \be\label{Eq:DegPol}
 \Pi=\mu_p/\mu_{100}.
 \ee
In the case of pair production, one could use the exact same kind of analysis as the one presented in Eqs.~\ref{Eq:ComptCounts}-\ref{Eq:DegPol} to determine the degree and angle of polarization from the azimuthal distribution of the electron-positron plane. I discuss in more detail the pair production polarimetry in Sec.~\ref{SSec:PairProdPol}.

The sensitivity of a polarimeter is usually described in terms of the minimum detectable polarization (MDP). For a detection of a given significance level, quantified as a $n_{\sigma}$ multiple of the Gaussian sigma,  MDP can be computed as following~\citep{Weisskopf:2010}:
\be\label{Eq:MDP}
MDP=\frac{n_{\sigma}}{\mu_{100}S_fA_{eff}}\sqrt{\frac{(S_fA_{eff}+B_r)}{T}},
\ee
where $B_r$ is the background counting rate, $S_f$ the source flux, $A_{eff}$ the effective area of the detector, and $T$ the observation time. The effective area $A_{eff}$ can be expressed in terms of the detection area $A$ and the efficiency of the detector, $\epsilon$\footnote{Not to be confused with $\epsilon(\theta)$ introduced in Eq.~\ref{Eq:RatioE}} as: $A_{eff}=\epsilon A$. Eq.~\ref{Eq:MDP} shows that, as expected, in order to achieve a lower minimum detectable polarization for a given source and for a given observation time one needs to have a detector with a higher response, i.e. higher $\mu_{100}$. However, as shown above, the polarization asymmetry decreases as $2m_e/E$ for $E\gg m_e$. Accordingly, a polarimeter based on Compton scattering becomes inefficient at energies of a few $\unit{MeV}$. Moreover, as one can see from Fig.~\ref{Fig:Regions}, in that energy range, for a moderate Z scattering element, $e^+e^-$ pair creation becomes dominant. The measurement of polarization in this high energy gamma-ray regime can be achieved by detecting the asymmetry in the azimuthal distribution of the electron-positron pair plane \citep[e.g.][]{Wick:1951}. However, this technique has severe limitations due to multiple Coulomb scatterings in the detector, and there are currently no successful polarization measurements for astrophysical sources in this regime. I will present in more depth pair production polarimeter designs and their inherent challenges in Sec.~\ref{SSec:Proposals}.

\subsection{Pair Production Polarimetry: Basic Principles}\label{SSec:PairProdPol}

For energies above the pair creation threshold, a photon can convert into a positron electron pair in the presence of the electric field of electrons or nuclei. This process is called triplet conversion and nuclear pair conversion, respectively. The use of nuclear pair conversion for polarimetry has been noted since 1950s~\citep[e.g.][]{Berlin:1950,Wick:1951}. Although almost seven decades have passed since then, only very recently linear polarization of photons was successfully measured using pair production in laboratory experiments~\citep[two such examples are][]{deJager:2007,Gros:2018}. No such measurements have yet been achieved in space for gamma rays of astrophysical origin. This slow advance is due to the confluence of two competing effects: a) the small cross section of pair production leads to a requirement of more material that serves as a convertor, and b) in order to achieve a high enough angular resolution, necessary for polarimetry, one needs less intervening material between the electron-positron pair and the detector in order to limit multiple scatterings. This, combined with the relatively low flux of high energy gamma rays of astrophysical origin, are the main reasons for the aforementioned ``$\unit{MeV}$ sensitivity gap.'' In Sec.~\ref{SSec:Proposals} I will explain how recent technological advances have lead to a large number of proposals for spacebourne instruments that should be capable of gamma ray astronomy, including polarimetry, at energies above the pair production threshold.

\begin{figure*}
 \begin{center}$
 \begin{array}{c}
 \includegraphics[scale=0.5]{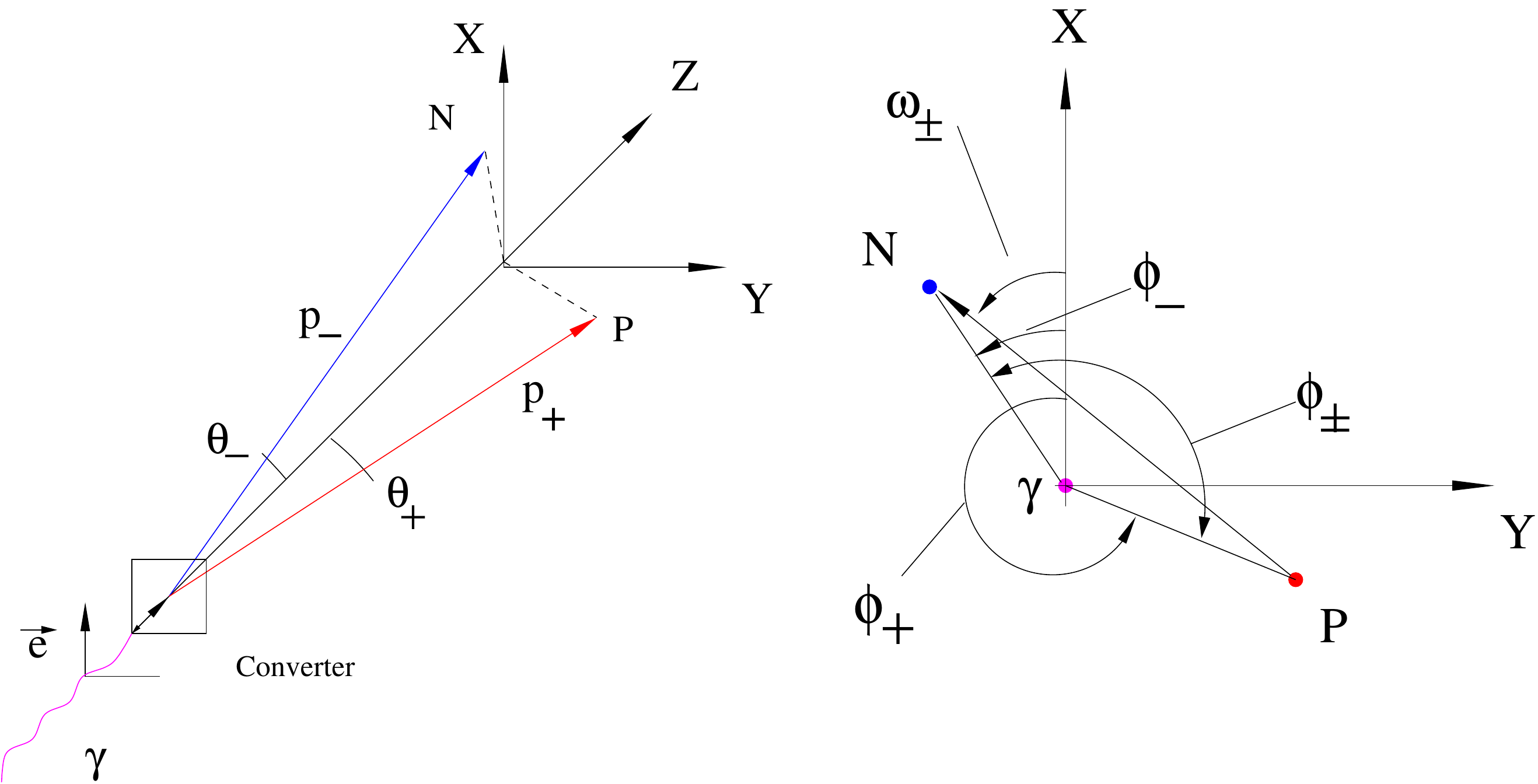}
 \end{array}$
 \end{center}
 \caption{The kinematics of the $e^+e^-$ pair production (left) and the azimuthal angles in the detector plane(right). $\phi_{\pm}$ is the coplanarity angle, $\omega_{\pm}$ is the angle between the polarization plane of the incident photon (Z-X plane in the figure) and the direction given by the P and N points which denote the positions of the crossing of the detector plane by the positron end electron, respectively. This figure is reproduced from~\cite{Wojtsekhowski:2003} with the permission from Elsevier.}
 \label{Fig:Kinempp}
 \end{figure*}

The main signature of linear polarized photons in the case of pair production is the asymmetry of the azimuthal distribution of the electron positron pair. For example, the case of nuclear pair conversion and when nuclear recoil is small, i.e. for high Z, the electron positron and incident photon momenta are nearly coplanar~\citep{Maximon:1962}, with the azimuthal orientation of the electron positron plane tending to align to the direction electric field vector, i.e. with the direction of polarization of the incident photon. In the case of triplet conversion, the recoil electron tends to be emitted in a plane orthogonal to the direction of polarization~\citep[e.g.][]{Boldyshev:1994}. Both of those asymmetries are experimental signatures of linearly polarized incident gamma-rays in a pair converter and could be used, in principle, for polarimetry. In general, the 1D differential interaction rate exhibits an azimuthal modulation that can be parametrized by the polarization asymmetry of the conversion process, $\mathcal{A}$~\citep[e.g.][]{Gros:2017}:

\be\label{Eq:DIR}
\frac{d\Gamma}{d\phi}\propto (1+\mathcal{A}P\cos2(\phi-\phi_0)),
\ee
where $P$ is the polarization fraction of incident photons, the angle $\phi_0$ gives its orientation, and $\phi$ is the azimuthal angle of the event, which essentially defines the orientation of the conversion event in a plane orthogonal to the direction of the incident photon. Note that there is no unique definition of this angle, and any shift in $\phi$ would lead to a change in the polarization asymmetry, $\mathcal{A}.$ Due to the azimuthal modulation in the interaction rate, the angular distribution of the electron positron plane can be fitted with a function that depends on the cosine of the azimuth angle, just as the one we used in Eq.~\ref{Eq:ComptCounts} the case of Compton polarimeters. When such a fit is used to determine polarization, one can estimate its RMS resolution in terms of the asymmetry factor $\mathcal{A}$:
\be\label{Eq:SigmaP}
\sigma_P=\frac{\sigma_{\mathcal{A}P}}{\mathcal{A}}\approx\frac{1}{\mathcal{A}}\sqrt{\frac{2}{N}},
\ee
with N being the number of points fitted. This shows directly that any dilution in the asymmetry factor leads to a loss in the precision of the measurement. The experimental determination of the asymmetry factor depends on the choice of the azimuthal angle, as I discuss further in the next paragraph. Theoretically, the polarization asymmetry can be estimated by using the full 5D analytic form of the scattering cross section obtained in the first order Born approximation by ~\citet{Berlin:1950,May:1951} and integrating over the polar and energy fraction final  state variables. This leads to a 1D expression similar to Eq.~\ref{Eq:DIR}. Using this technique,~\citet{Gros:2017} has obtained a value of $\pi/4$ for the low energy asymptote for the polarization asymmetry. In the high energy limit $\mathcal{A}\approx 1/7$, as obtained by ~\citet{Boldyshev:1972}. In the intermediate regime, the theoretical value of the polarization asymmetry is approximately 0.2. Its important to note that all of those values are valid when the azimuthal angle $\phi$ is chosen to be equal to the bisector angle of the electron positron pair: $\phi\equiv(\phi_++\phi_-)/2$. For the geometry of the conversion event see Fig.~\ref{Fig:Kinempp}.

 %%%%%%%%%% Figure Kinem PP %%%%%%%%%%%%%%%

From an experimental perspective, at high energies nuclear pair production dominates and the azimuthal angle is measured from the $e^+e^-$ pair. As  noted earlier, there is no unique definition of the azimuthal angle, as one could use various final state azimuthal angle variables to quantify the orientation of the final state with respect to the direction of polarization of the incident photon. For example, most of the analytical expressions use the bisector angle of the electron positron pair: $\phi=(\phi_++\phi_-)/2$. For triplet conversion, it might be more convenient to use the azimuthal angle of the recoiling particle ($\phi_r$) in the field of which the pair conversion takes place, since this becomes a measurable quantity. Another choice, favored by experimentalists, is the pair plane azimuthal angle $\omega_\pm$ (see Fig.~\ref{Fig:Kinempp}). No matter what definition one choses for the azimuthal angle, there are two approaches to determining the polarization fraction $P$. One can either fit the distribution of events as a function of the polarization angle using the cosine modulation of Eq.~\ref{Eq:DIR}, as discussed above, or use a moments method. The moments method is based on an appropriate choice of statistical weights $w(\phi)$ for each event~\footnote{The discussion of the weights method follows ~\citet{Bernard:2013, Gros:2017}, and the interested reader is encouraged to consult those references for more details.}. A judicious choice of $w(\phi)$ allows one to extract an estimator for observable parameters of interest from the measured average:~$\langle w \rangle$. Given the expected distribution from Eq.~\ref{Eq:DIR}, the expectation value of $w(\phi)$ is:

\be\label{Eq:Expectation}
\langle w\rangle=\int\frac{w(\phi)}{\Gamma}\frac{d\Gamma}{d\phi}d\phi.
\ee

If the direction of the polarization of the incident photon ($\phi_0$) is unknown, which is generally the case, then one needs the average over the distribution of two weight functions: $\langle\cos2\phi\rangle$ and $\langle\sin2\phi\rangle$ to extract the two unknown parameters of the azimuthal distribution: $\mathcal{A}P$ and $\phi_0$. Using Eqs.~\ref{Eq:DIR} and ~\ref{Eq:Expectation} it is straightforward to show that the quantities of interest become:

\begin{eqnarray}\label{Eq:APPhi}
	\mathcal{A}P&=&2\sqrt{\langle\cos2\phi\rangle^2+\langle\sin2\phi\rangle^2}\\
	\phi_0&=&\frac{1}{2}\arctan\left(\frac{\langle\sin2\phi\rangle}{\langle\cos2\phi\rangle}\right)
\end{eqnarray}

Propagating errors leads to the following expressions of the statistical uncertainty of $\mathcal{A}P$ and $\phi_0$ in terms of the  average ($\langle w\rangle$) and the variance ($\sigma(w)\equiv R.M.S.(w)/\sqrt{N}$) of the two weight functions:

\begin{eqnarray}\label{Eq:UncAPPhi}
\sigma_{\mathcal{A}P}&=&2\frac{\avgc\sigma(\cos2\phi)+\avgs\sigma(\sin2\phi)}{\sqrt{\langle\cos2\phi\rangle^2+\langle\sin2\phi\rangle^2}}\\
\sigma_{\phi_0}&=&\frac{1}{2}\frac{\avgc\sigma(\sin2\phi)+\avgs\sigma(\cos2\phi)}{\langle\cos2\phi\rangle^2+\langle\sin2\phi\rangle^2}
\end{eqnarray}

For calibration purposes, one uses a beam of known polarization $P$ and energy (or an event generator~\footnote{For benchmarks of various available event generators see ~\citet{Gros:2017b}}) in order to determine the asymmetry factor $\mathcal{A}$, whereas for polarimetry the analyzing power ($\mathcal{A}$) is known from calibration and the degree of linear polarization can be determined. When the direction of the polarization of the incident photon is known, we only need one statistical weight function, since we have only one unknown, $\mathcal{A}P$. Taking $w$ to be $2\cos2\phi$ leads to $\mathcal{A}P=2\avgc$ and $\sigma_{\mathcal{A}P}=\sigma(2\cos2\phi)$. Using the definition $\sigma(w)=\sqrt{\frac{\langle w^2\rangle-\langle w\rangle^2}{N}}$ we arrive at the following expression for the uncertainty in the polarization measurement:

\be
\sigma_{P}=\frac{1}{\mathcal{A}P\sqrt{N}}\sqrt{2-(\mathcal{A}P)^2}.
\ee

Comparing this result to the uncertainty when the determination of the polarization is made based on the fit of the azimuthal distribution of the number of events (Eq.~\ref{Eq:SigmaP}) we note that the former is always smaller. Also note that in the limit of small asymmetry and/or polarization one recovers the result in Eq.~\ref{Eq:SigmaP}. Moreover, one can see once again the importance of choosing the azimuthal angle $\phi$ variable in such a way to maximize the analyzing power $\mathcal{A}$. There are two limitations here. First, for the case of nuclear conversion, the azimuthal angle $\phi$ cannot be measured directly, in view of the recoil of the nucleus going undetected. Using the azimuthal angle of one of the tracks (e.g. $\phi_+$ or $\phi_-$ in Fig.~\ref{Fig:Kinempp} for the positron/electron tracks) leads to a decrease in the effective asymmetry, especially at energies below $100\unit{MeV}$ ~\citep[e.g.][]{Bernard:2013}. A partial recovery in sensitivity could be achieved if one uses the angle between the polarization plane of the incident photon and direction that connects the positron and the electron in the detector plane ($\omega_{\pm}$ in Fig.~\ref{Fig:Kinempp}), as shown for example by ~\citet{Yadigaroglu:1997}. Recently ~\citet{Gros:2017} showed that the use of the bisector angle of the lepton pair ($(\phi_++\phi_-)/2$) leads to a maximization of the obtained experimental value of the polarization asymmetry, and therefore to a better precision of the polarization measurement. The second limitation comes from multiple Coulomb scatterings of the charged tracks, leading to an exponential suppression of the asymmetry factor: $\mathcal{A}_{eff}=e^{-2\sigma_{\phi}^2}\times\mathcal{A}$,
% \be\label{Eq:ExpSup}
% \mathcal{A}_{eff}=e^{-2\sigma_{\phi}^2}\times\mathcal{A},
% \ee
with $\sigma_{\phi}$ being the azimuthal angular resolution. It can be approximated in terms of the radiation lengths, L, traversed in the material: $\sigma_{\phi}\simeq 14\sqrt{L}$~\citep{Kotov:1988,Mattox:1990}. This parametrizes the degradation of the precision due to experimental effects, in this case multiple scatterings. As $\mathcal{A}_{eff}$ decreases the minimum degree of polarization is quickly degraded, as is the precision of the measurement (see Eq.~\ref{Eq:SigmaP}).

In a laboratory setting, in the case of nuclear pair production and using the Laser Electron Photon (LEP) beam line at SPring-8,  the asymmetry in the azimuthal distribution of the electron positron plane has been observed~\citep{deJager:2007}. This thus demonstrates that gamma-ray polarimetry with pair production is possible. Due to thick converter foils EGRET, AGILE, and GLAST have a suppression factor of the order of $10^{-4}$ and are therefore insensitive to polarization. This is one of the main limitations of pair production telescopes as polarimeters, as one needs to be able to define the plane electron positron pair by sampling very few radiation lengths (RLs). For a slab detector, made by a succession converter slabs interweaved with tracking detectors in which electrons are tracked, the dilution of the analyzing power is too severe. Thus slab detectors are found to be unfeasible~\citep{Bernard:2013}. The solution is to consider active target technology, in which the conversion and detection is done by the same device. In Sec.~\ref{SSec:Proposals} I will discuss the latest proposals using this approach. 

As mentioned before, one can increase the polarization asymmetry by using optimal kinematic variables, or judicious event selection~\citep[e.g.][]{Olsen:1959,Maximon:1962,Endo:1993,Bakmaev:2008}. This, and a presentation of experimental effects that affect polarization measurements at energies above the pair production threshold, is discussed in detail in~\citet{Bernard:2013, Gros:2017}. For example~\citet{Bernard:2013} found that the gain in asymmetry using the event selection strategy is more or less cancelled by the reduced sample size, which affects the statistics, and thus the analyzing power ($\mathcal{A}$). In addition~\citet{Bernard:2013} shows that the measurements are badly affected by multiple scattering. A novel approach, based on event weighting to extract the asymmetries, has been introduced recently by~\citet{Pretz:2018}. They show that the resulting estimator has advantages over those previously considered in the literature.

% An alternative would be the use of thin, active targets, where the conversion and tracking are performed by the same device.
% %%%%%Here insert ref from Jager 2007!!

%
%
\section{Past and future Hard X-ray and gamma-ray polarimeters}\label{Sec:Instruments}
In this section I continue the discussion of instruments that were already used as gamma-ray polarimeters (Sec.~\ref{SSec:Past}) and review some of the most important proposals and the various experimental advances for such instruments that are currently at various stages of development (Sec.~\ref{SSec:Proposals}). Most of the latter are based on pair-production polarimetry, the basics of which were discussed in Sec.~\ref{SSec:PairProdPol}. The challenges of the field, making it an extremely delicate endeavor, are due to the fact that the polarized component constitutes only an small component of the total radiation emitted by a given source.  Therefore, best detection is achieved only for the brightest sources. Existing instruments, such as IBIS and SPI (on INTEGRAL) have been used successfully for hard X-ray and soft gamma-ray polarimetry, as discussed in Sec.~\ref{SS:Challenges}. Results are promising, but inconclusive at this stage. This is in contrast to the situation in the optical and radio bands. For the former, information regarding light intensity is available. At radio wavelengths, one can record both amplitude and phases for electromagnetic signals from astrophysical sources of interest, making wide field spectro-polarimetric surveys such as GMIMS (The Global Magneto-Ionic Medium Survey) possible. For a more detailed perspective, see~\citet{Hajdas:2010} and~\citet{Bernard:2013}, two recent reviews focusing on high and medium energy gamma-ray polarimetry.
\subsection{Past and current hard X-ray and gamma-ray polarimeters for astrophysics}\label{SSec:Past}

Despite being an extremely challenging endeavor, gamma-ray polarimetry at energies above $\sim 100\unit{keV}$ has progressed significantly in the past 20 years. Below I discuss some of the most relevant results and their implications. The large area imaging Compton Telescope COMPTEL and the Burst and Transient Source Experiment (BATSE) were both on board of the NASA Compton Gamma Ray Observatory (CGRO). This was a satellite that carried four instruments capable of exploring the electromagnetic spectrum in six decades of energy, from 30 keV to $30 \unit{GeV}$ which was operational between 1991 and 2000. COMPTEL operated in the $0.75$ to $30$ MeV range with a FOV of $1 \unit{sr}$ and an $A_{eff} \lesssim 20 \unit{cm^2}$.  However, the instrument was not optimized for polarimetry, and serious systematic effects and low statistics made the analysis of polarimetric observations impossible, even for the Crab nebula. COMPTEL's failure as a polarimeter highlighted the importance of instrument geometry, which needs to be optimized to maximize the acceptance of photons Compton scattered at high angles, where the polarimetric modulation is greatest. BATSE was sensitive to photons with energies between $40$ and $600 \unit{keV}$ and had a full $4\pi \unit{sr}$ FOV. Out of the $3000$ total GRBs in the BATSE catalogue polarimetric analysis was possible only for two of them, with lower limits on the polarization of $\Pi>30\%$ and $50\%$ respectively \citep{Willis:2005}.

The Reuven Ramaty High Energy Solar Spectroscopic Imager (RHESSI) was launched in 2002 \citep{RHESSI:2002} as a NASA-SMEX mission for imaging the Sun at energies between $3$ keV and $20$ MeV. It was also capable of measuring polarization of hard X-rays \citep{RHESSI:2002Pol} by using Be scattering element located in the cryostat that houses the nine germanium detectors. The low energy photons reach the detectors after being scattered off the Be element will have an azimuthal distribution that carries the information regarding the linear polarization of the incident photons. At higher energies, polarization can be extracted for gamma-rays up to 2 MeV from photons scattered between neighboring Ge detectors. In 2003 \citet{Coburn:2003} reported, using data from RHESSI, very high ($\Pi\simeq 80\%$) polarization levels from GRB021206, a result that lead to a wave of excitement and accelerated the progress of hard X-ray and gamma-ray polarimetry. Moreover, RHESSI was successfully used to study the polarization of numerous solar flares \citep[e.g.][]{McConnell:2003, Suarez-Garcia:2006, Boggs:2006, Emslie:2008}.

 In 2002 the European Space Agency (ESA) launched the INTErnational Gamma-Ray Astrophysics Laboratory (INTEGRAL) mission, with the aim of ``providing a new insight into the most violent and exotic objects of the Universe, such as black holes, neutron stars, active galactic nuclei and supernovae~\footnote{\url{https://heasarc.gsfc.nasa.gov/docs/integral/integralgof.html}}.'' The instrument was designed to fill the sensitivity gap between traditional X-ray telescopes and space and ground experiments that concentrate on high and very high gamma-ray astronomy. Data from INTEGRAL is also being used in the study of processes such as the formation of new chemical elements and GRBs - the most energetic phenomena in the Universe. As previously mentioned, two of its main instruments, the IBIS imager and the SPI spectrometer, have polarization detection capability, even if they were not specifically designed with that purpose \citep{Lei:1997,Forot:2007}.

SPI can detect photons with energies between $20~\unit{keV}$ and $8~\unit{MeV}$ and is sensitive to polarization, as the detector plan consists of 19 independent Ge crystals that can operate as a polarimeter since the anisotropy characteristics of the Compton diffusions can be used to extract information regarding the polarization of the incident gamma-rays~\citep[see][]{Chauvin:2013SPI}. Data from SPI was used to determine the polarization of a few bright GRBs. Independent analysis by two groups confirmed a very high polarization degree of the gamma-rays from GRB 041219a \citep{McGlynn:2007,Kalemci:2006}. This detection opened an observational window that provides additional information regarding the emission mechanisms of GRBs, which is still an unresolved and critical issue in astrophysics! SPI was used by ~\citet{McGlynn:2009} to place an upper bound of $60\%$ on the polarization fraction of GRB 061122. Data of the polarization from the Crab nebula at gamma-ray energies taken with SPI \citep{Dean:2008} and IBIS \citep{Forot:2008} show that at energies higher than 200 keV the flux is highly polarized ($\sim50\%$). Furthermore, the position angle is consistent with the  projection on the sky of the spin axis of the central neutron star. This alignment was employed by \citet{Maccione:2008} to place constraints on Lorentz Invariance Violation (LIV) in QED by using the generic prediction of vacuum birefringence present in LIV theories. This would imply that the direction of polarization rotates during propagation, as one of the effects of LIV is to induce terms that make dispersion relation different for left and right handed photons, and therefore different propagation speeds, which leads to a rotation of the polarization angle. See Sec.~\ref{Sec:LIV} for more details. As this is a propagation effect, one can get tighter constraints on LIV with more distant polarized gamma-ray sources. For example,~\citet{Stecker:2011} used the polarization level of GRB041219a, as determined by~\citet{McGlynn:2007} using the INTEGRAL/SPI instrument to determine, the most stringent, at the time, constraint on the parameter that measures deviation from Lorentz Invariance, $|\xi|\lesssim 2.4\times 10^{-15}$. In 2013, one of the best contemporary constrains on LIV models ($|\xi|\lesssim 3.4\times 10^{-16}$) was been obtained by \citet{Gotz:2013GRB} using INTEGRAL-IBIS data to estimate the polarization of the GRB061122 emission in the X-ray and soft gamma-ray energy bands.

The Imager IBIS on board the INTEGRAL satellite was used in its Compton mode to study the polarization level for gamma-rays from several GRB's: e.g. GRB 041219A~\citep{Gotz:2009GRB}, GRB 061122~\citep{Gotz:2013GRB}, GRB 120711A \citep{Gotz:2013GRBConf}. In 2014,~\citet{Gotz:2014GRB} reported finding a polarization fraction $\Pi>28\%$ at a $90\%$ CL in the prompt emission of GRB 140206A. Using optical spectroscopy data from the Telscopio Nazionale Galileo (TNG), the distance to this GRB was determined to correspond to a redshift $z\sim 2.74$, making GRB 14020A the most distant Gamma-Ray Burst for which polarization measurements have ever been observed! This large distance and the polarization measurement with IBIS leads to an improvement of the previous constraints on LIV by a factor of three, i.e.  $|\xi|\lesssim 1\times 10^{-16}$. The Japanese Gamma-Ray burst polarimeter GAP on the power sail platform IKAROS confirmed detection of polarization in three bright GRB's: 100826A~\citep{Yonetoku:2011}, 110301A, and 110721A~\citep{Yonetoku:2012}. Even if not extremely statistically significant yet, as all of these detections are at the $3-4 ~\sigma$ level. Taken together, GRB polarization independent results from IBIS, SPI and GAP indicate a high level of polarization should be present in the prompt emission of GRBs. This has important theoretical implications on the emission mechanism and the magnetic composition of gamma-ray bursts. Another important astrophysical object for which polarization was studied with IBIS is the galactic black hole  (BH) Cygnus X-1. \citet{Laurent:2011} and~\citet{Jourdain:2012} report finding strong polarization at energies above $400~\unit{keV}$ using INTEGRAL/IBIS or INTEGRAL/SPI data, respectively. This would indicate that the gamma-ray emission from the Cygnus X-1 BH binary is probably due to the jet previously detected in the radio band, a conclusion that is firmly reinforced by ~\citet{Rodriguez:2015}.

In September 2015, the Indian AstroStat mission was launched on an IRS-class satellite into a near Earth equatorial orbit. The Cadmium-Zinc-Telluride (CZT) Imager  is an instrument onboard AstroStat that is also capable of polarization studies~\citep{Vadawale:2015}, having being calibrated on the ground for hard X-ray polarimetry. In fact, CZT has detected, on its very first day of operation a long duration, relatively faint, gamma ray burst that shows hints of polarization: GRB 151006A. In a recent study~\citep{Vadawale:2018}, data from CZT was used to determine the most accurate to date polarization degree of the hard X-ray emission from the Crab pulsar nebula, confirming the early indication of a strongly polarized off pulse emission. Intriguingly, the data seems to suggest a variation of the polarization properties within the off-pulse region, which could not be explained by current theoretical models of pulsar emission. However, this behavior was not confirmed by measurements in an partially overlapping energy band by PoGo+~\citep{Chauvin:2018}. This is a very interesting open debate, with potential profound implications for our understanding of pulsar emission mechanism. Future, more accurate measurements covering a larger range of energies are needed to settle this.

An excellent review of the numerous recent advances in the field of hard X-ray polarimetry and the various detector technologies that made this possible can be found in~\citet{Fabiani:2018}. In contrast, progress in the field of gamma-ray polarimetry above the pair creation threshold was relatively slow and limited by position resolution, the number of readout channels, and multiple Coulomb scatterings in the detector. Up to the present, all gamma-ray telescopes in the $\unit{MeV}-\unit{GeV}$ regime suffer from a strong degradation of the angular resolution at low energy, thus making background rejection extremely difficult. In the case of GRBs the background is not an issue, and this explains in part why the current polarization measurements for gamma-rays of astrophysical origin are mostly restricted to GRBs. However, numerous other cosmic sources that have potentially polarized emission, such as pulsars (curvature radiation), Pulsar Wind Nebula (synchrotron radiation), Active Galactic Nuclei (synchrotron radiation or Inverse Compton), etc. All of them have their peak emissivity in the $\unit{MeV}-\unit{GeV}$ range. It is therefore essential for the understanding of high energy astrophysics to have gamma-ray data (both spectroscopic and polarimetric) that bridges the sensitivity gap discussed before (see Fig.~\ref{Fig:SensGap}).  The next generation of Compton telescopes would use the lessons learned from CGRPO/COMPTEL and improve the sensitivity in the $200\unit{keV}-50\unit{MeV}$. At higher energies the sensitivity of Compton based polarimeters quickly degrades, leading to the need for a new approach: pair production polarimeters. In Sec.~\ref{SSec:PairProdPol}, I presented the basic principles of pair production polarimetry. In the next sub-section, I discuss several of the most important proposals and the respective techniques that attempt to overcome the inherent experimental challenges that hindered the advance of medium and high energy gamma-ray polarimetry so far.

\subsection{Proposals and instruments being built}\label{SSec:Proposals}

The need for dedicated polarimeters in the MeV-GeV regime has long been recognized, and, as noted earlier, the theory that would in principle allow one to extract polarization measurements from pair production events has been developed in the early 1950s. Due to practical limitations, discussed in the previous section, the implementation of such techniques was relatively slow. However, in the past decade or so, a number of technological advances have lead to an increased interest in this field, and multiple gamma-ray telescopes that could be used in the near future for medium and high energy polarimetry have been suggested. In this section I will present a non-exhaustive list, accompanied by brief descriptions, of the recent proposals for such instruments. Relevant detector technologies and their common abbreviations used in the literature include: Time Projection Chambers (TPC), silicon trackers (Si), Cadmium zinc telluride (CdZnTe or CZT) detectors, cesium iodine scintillators (CsI), and scintillating fibers (fib.), just to name a few.

\begin{itemize}

	\item{\bf{AMEGO}} The All-sky Medium Energy Gamma-ray Observatory (AMEGO) is a proposal for a NASA Probe class mission that aims to improve on the continuum sensitivity of previous gamma-ray detectors by a factor of $20-50$~\citep{AMEGO:2017}. Moreover, AMEGO should be sensitive to linear polarization. For example, it is estimated that in the Compton mode, for a source with an intensity $1\%$ of the Crab, a $10^6s$ exposure would lead to a minimum detectable polarization (MDP) of $20\%$. In order to achieve those goals, AMEGO will operate as a Compton telescope at energies below $10~\unit{MeV}$ and as a pair production telescope at higher energies. In the Compton regime, its use of solid state technology and compact geometry should lead to a substantial improvement in sensitivity relative to COMPTEL. The instrument has four main subsystems: a double-sided silicon detector (DSSD) tracker, a CdZnTe calorimeter, a CsI(Tl) calorimeter, and a plastic scintillator anti-coincidence detector (ACD). For further details, please see ~\citet{AMEGO:2017}. AMEGO is based on studies for ComPair, GRIPS, and MEGA. For more details on each of those projects, see the list below.  

	\item{\bf{All-sky-ASTROGAM}} The ASTROGAM mission concept~\citep{DeAngelis:2016} was proposed to ESA as part of the M4 call for Missions. Unfortunately, it was not selected for a mission study. As a result, the enhanced ASTROGAM (e-ASTROGAM) version of the mission concept was prepared~\citep{DeAngelis:2017} and  submitted as part of the M5 Call for Missions. Its detector includes 56 layers of $10\times 10$ double sided thin silicon strip detectors (DSSDs) which lead to a very good angular resolution  and will allow for groundbreaking polarimetry capabilities over the entire energy range, both in Compton and pair production regimes. e-ASTROGAM is dedicated to the study of the non-thermal universe, being sensitive to gamma rays in the $0.3\unit{MeV}-3\unit{GeV}$ range. A combination polarimetric and spectroscopic observations from e-ASTROGAM could lead to the elucidation of the nature and impact of relativistic flows from the most energetic Galactic and Extra-Galactic sources. The anticipated performance of e-ASTROGAM as a polarimeter in the low energy ($<5\unit{MeV}$) Compton regime was estimated, based on simulations with the MEGALib software and results were published in ~\citet{Tatischeff:2017}. The results indicate that e-ASTROGAM should be able to perform unprecedented polarization measurements of $\sim\unit{MeV}$ gamma rays. For the pair production regime, performance evaluation based on simulations is still in progress, but preliminary estimates are encouraging. Unfortunately, as was the case for the ASTROGRAM proposal, the e-ASTROGAM mission concept was not selected for funding  by ESA either. Despite all of those setbacks, a third attempt is in progress, in the form of the All-Sky-ASTROGRAM proposal. For details please see~\citet{Tatischeff:2019}. 

	\item {\bf{APT/AdEPT}} The advanced pair telescope (APT) concept presented by~\citet{Bloser:2004Concept} is  an imaging gamma-ray polarimeter operating from $\sim 50$ MeV to $\sim$ 1 GeV. It uses pixelized gas micro-well detectors to record the electron-positron tracks from pair-production events in a large gaseous Time Projection Chamber (TPC). The polarization sensitivity of APT was estimated using Geant4 Monte Carlo simulations.  Results were  preliminary  as the  simulation has been shown to produce electron-positron multiple scattering results with an rms scattering angle $\sim$ 20\% lower than that predicted by Moli\'{e}re theory.  Though better statistics are needed in order to improve the determination of the polarization angle,  those preliminary results suggested that this APT concept would have a useful polarization sensitivity for bright sources around 100 MeV~\citep{Bloser:2004Concept}. \citet{Hunter:2013} describe the design of the Advanced Energetic Pair Telescope (AdEPT), a pair production telescope dedicated to medium energy gamma-ray polarimetry. AdEPT, an improved design based on APT, would have better sensitivity than FERMI-LAT up to $\sim200\unit{MeV}$. This instrument aims to achieve a high angular resolution and low minimum detectable polarization (MDP) by using a gaseous medium to act as a continuous tracking detector, and thus allow for a full reconstruction of the electron and positron tracks from pair production. This proposed design, based on the Three-Dimensional Track Imager (3-DTI) technology, is in contrast with previous pair production polarimeters that relied on electron tracking hodoscopes that consist of tracking detectors interleaved with metal foils that serve as converters. The authors estimate that AdEPT would be able to achieve a MDP of $\sim 10\%$ for a $10 \unit{mCrab}$ source in $10^6\unit{s}$.

	\item {\bf{ComPair}} The Compton-Pair Production (ComPair) space telescope is a proposed mission to investigate gamma rays in the $200\unit{keV}-500\unit{MeV}$ energy range~\citep{Moiseev:2015}. Its design focuses on high energy and angular resolution and should achieve sensitivities larger by a factor of $20-50$ when compared to COMPTEL. It will utilize Si-strip and CdZnTe-strip detector technology, and photon polarization measurements in space should be possible in view of Si-strip tracker planes that act as both converter and tracker. Even if not designed primarily as a polarimeter, it is estimated that for the  Crab system polarization of $20\%$ and higher should be detectable in one month of exposure.

	\item {\bf{DUAL}} The DUAL Gamma-Ray mission \citep[e.g.][]{Knodlseder:2005GRI,Boggs:2010GRI} is a proposed mission for gamma-ray astronomy, focusing on deep observations of SNe Ia and wide field nuclear gamma-ray astrophysics, made possible by the recent developments of compact Compton telescopes and Laue lenses. Its target are soft gamma-ray photons, with energies in the $50\unit{keV}-10\unit{MeV}$ range. Thus it will overlap significantly with IBIS and SPI. One of DUAL's mission instruments, the Wide-Field Compton Telescope (WCT) will be used for medium-sensitivity large-scale exposures. For very deep, pointed observations of selected narrow-field targets the Laue-Lens Telescope (LLT) in combination with the WTC as a focal plane. Polarimetry with DUAL/LLT should be possible, since a Compton camera is inherently sensitive to gamma-ray polarization.

	\item {\bf{GRAPE}} The Gamma Ray Polarimeter Experiment (GRAPE)  is a large FoV instrument sensitive to polarization of gamma rays in the $50-500~\unit{keV}$ band. Its primary mission is the study of GRBs over the entire sky \citep{Bloser:2005Grape}; however, the instrument can also be used for point sources. It had its first science flight in September 2011, having being launched from Ft. Sumner, NM on a balloon platform. Polarization sensitivity is limited by several factors, but during the 26 hours flight it gathered sufficient data to place upper limits on the soft gamma-ray polarization for the Crab nebula and two M-Class solar flares. A second successful balloon flight, in 2014, coupled to several design efforts focused on orbital payloads, lead to an improvement in the GRAPE polarimeter concept and there is hope that in the near future GRAPE will be flown on a long duration balloon platform in order to collect polarimetry data on a large number of GRBs.

	\item{\bf{GRAINE}} The Gamma-Ray Astro-Imager with Nuclear Emulsion (GRAINE)~\citep{GRAINE:2015, GRAINE:2018} is a large aperture, high resolution ($0.08\deg at \sim1\unit{GeV}$), polarization sensitive emulsion telescope sensitive to cosmic gamma-rays in the $10\unit{MeV}-100\unit{GeV}$. This instrument is based on nuclear emulsion technology, which allows a precise determination of the electron positron angles at the conversion vertex and the azimuthal angle of the pair plane. As such, GRAINE has an angular resolution about one order of magnitude higher than Fermi-LAT, and thus should be capable of polarimetry. The first GRAINE balloon experiment took place in 2011, demonstrating the feasibility of the emulsion based gamma-ray telescope concept. In 2015, a second balloon flight took place, and data from that experiment was analyzed and reported by~\citet{GRAINE:2018}. By using an automated selection process the GRAINE team was able to record approximately $10^6$ gamma-ray events! In addition, the experiment confirmed the excellent angular resolution expected. Currently there are plans for a third balloon experiment with the aim of detecting sources of cosmic gamma-rays. Using the SPring-8/LEPS facility,~\citet{Ozaki:2016} has demonstrated the feasibility the polarimeter that will be used by GRAINE, with a detection of a non-zero modulation in the azimuthal angle of the electron-positron pair at the $\sim 3\sigma$ level being reported.  

	\item{\bf{GRIPS}} In 2011 the Gamma-Ray Imaging, Polarimetry and Spectroscopy (GRIPS) was proposed by~\citet{Graeiner:2012} as a continuously scanning all-sky survey from $200 \unit{keV}$ to $80\unit{MeV}$. It will consist of three ESA instruments: the Gamma-Ray Monitor (GRM), the X-Ray Monitor (XRM), and the InfraRed Telescope (IRT). Since the target energy sensitivity range includes the crossover energy between Compton scattering processes and pair-production ($\sim 8\unit{MeV}$ for typical detectors), the GRM will employ two separate detectors, one for the Compton regime and one for the pair production regime.  The aim of GRIPS is to analyze astrophysical sources of gamma-rays such as GRBs, blazars, SN explosions, the source of positrons in our galaxy, nucleosynthesis, extreme particle accelerators in the universe (e.g. pulsars, magetars), and the radiation they produce. Polarimetry will play a central role in this mission. One of its aims is to ``decipher the mechanisms of jet formation in accreting high-spin black holes systems''~\citep{Graeiner:2012} such as GRBs and blazars by using measurements of  polarization of their gamma-ray emissions.

	\item{\bf{GAMMA-400}} The GAMMA-400 project, lead by the Lebedev Physical Institute \citep{Ginzburg:2007GAMMA400}, aims to obtain gamma-ray data in the $100\unit{MeV}-3000 \unit{GeV}$ range and is scheduled to be launched on the Russian space platform Navigator in 2019. Note that its energy range overlaps partly with that of the Fermi-LAT, extending it to $3~\unit{TeV}$. However, it will not be sensitive in the ``$\unit{Mev}$ gap.'' Its primary aim is the determination of the nature of Dark Matter(DM) in the Universe, by looking for gamma-rays or cosmic electrons and positrons that could be the product of DM annihilations of decay. Other important problems in astrophysics that could be solved with the help of this telescope include the nature of GRBs, the search and identification of new discrete gamma-ray sources of high energy, the measurement of energy spectra of galactic and extragalactic diffuse and isotropic gamma-radiation, etc. Even if not designed primarily as a polarimeter, GAMMA-400 data could have high enough angular resolution to be used to extract the linear polarization fraction $P$ for gamma-ray sources above the $e^+e^-$ pair creation threshold by using the azimuthal distribution of secondary particle momenta.

	\item{\bf{HARPO}} Time Projection Chambers (TPCs) are widely used and highly reliable particle detectors in high-energy particle physics experiments on the ground. Their basic principle of operation is to collect and precisely locate in a 2D anode plane ionization electrons produced by the passage of high-energy charged particles in a volume of gas subjected to an electric field. For a full 3D reconstruction the drift time of the ionization electrons is measured and a vertex finder can be used to identify conversion vertices and pseudo-tracks. The concept of a Time Projection Chamber (TPC) gaseous detector with high angular resolution and sensitive to polarization of gamma-rays of astrophysical origin with energies ranging between a few MeV to a few hundred MeVs has been recently considered in the literature~\citep[e.g.][]{Bernard:2012}. At energies of the order of $\sim\unit{GeV}$ and above, the photon flux is the main limitation. In order to be sensitive to polarization of such high energy photons, the TPC would need to become more voluminous and therefore heavy, limiting somewhat the energy range to which TPCs are optimal for use as spaceborne detectors. However, they are certainly very promising candidates in closing the $\unit{MeV}$ sensitivity gap (see Fig.~\ref{Fig:SensGap}) of gamma ray astronomy. Preliminary estimates of the spatial resolution show that TPCs could enable improvement up to an order of magnitude in single-photon angular resolution with respect to Fermi-LAT for $100\unit{MeV}$ photons~\citep{Bernard:2012}. HARPO (Hermetic ARgon POlarimeter), a demonstrator for this TPC concept, was built with the purpose of validating on the ground the performance of a TPC as a high energy polarimeter. In 2014, the HARPO TPC used the NewSUBARU polarized photon beam line for calibration purposes ~\citep{Gros:2018} at 13 distinct energies ranging between $1.74~\unit{MeV}$ and $74~\unit{MeV}$. One of the most significant results of that run is the measurement of the polarization asymmetry for pair production photons below $50\unit{MeV}$, demonstrating the potential of TPCs to be used as low to medium energy pair production polarimeters.

	\item The {{\bf{H}}igh {\bf{E}}nergy {\bf{P}}hoton {\bf{P}}olarimeter for {\bf{A}}strophysics} is a proposal by ~\citet{Eingorn:2018JATIS} for a dedicated space-borne polarimeter that would be sensitive to linear polarization of gamma-rays in the $10\unit{MeV}-800\unit{MeV}$ range, by using already proven silicon micro-strip detector (MSD) technology. This polarimeter is based on a concept demonstrator polarimeter, which was tested for the first time using the laser back-scattering $\unit{GeV}$ linearly polarized photon beam at Spring-8/LPS. For that run, an analyzing power very close to the theoretical limit of $20\%$ was observed~\citep{deJager:2007}. This was one of the first measurements of linear polarization of photons above the pair creation threshold, demonstrating the usefulness of the MSD technology for gamma-ray polarimetry. The space-borne polarimeter would consist of 30 cells, each including one  double sided MSD with 2D readout of 0.6 mm thickness and one double sided MDS with 2D readout of 0.3 mm thickness. Each of those MDS planes act as a convertor for pair production events. The main advantage of using this pair configuration is that it allows for an un-ambiguous determination of the geometric parameters of the conversion event.  Using Monte Carlo simulations, it was estimated that at $200\unit{MeV}$ the instrument will have an angular resolution of $\sim 5$ mrad.  In a one year long observation, it should achieve $\sim 6\%$ accuracy in polarization measurements for the Crab pulsar, for gamma rays below $200\unit{MeV}$. This polarimeter would therefore have the ability to measure polarization of gamma-rays in the un-explored $\unit{MeV}-\unit{GeV}$, and as such would constitute a significant advance for gamma ray astrophysics.

	\item {\bf{MEGA}} The Medium Energy Gamma-Ray Astronomy (MEGA) telescope has been proposed by Bloser and collaborators in a series of papers ~\citep{Bloser:2001MEGA,Kanbach:2004,Kanbach:2005,Bloser:2006MEGA,Bloser:2006MEGAb}. This  Compton telescope was designed to be the successor of COMPTEL and OSSE experiments on the Compton Gamma-Ray Observatory and planned to achieve a tenfold improve in sensitivity in the $0.5-50\unit{MeV}$ band. As an advanced Compton telescope with improved angular and spatial resolution, it would have had the capability, in principle, to be sensitive to polarization via the azimuthal modulation of the distribution of scattered photons~\citep[e.g.][]{McConnell:2004}. MEGA could operate in both Compton and pair production modes and its design was based on a stack of double sided Si strip detectors surrounded by a pixellated CsI calorimeter. The prototype instrument was calibrated in the laboratory and results of this experiment are presented in~\citet{Bloser:2006MEGA}. Although not materialized as a mission, the MEGA proposal is an important conceptual and practical advancement, as it was one of the first prototype designs targeted at covering the $\unit{MeV}$ sensitivity gap. Another, similar Compton scattering based concept instrument capable of gamma-ray polarimetry is the Advanced Compton Telescope (ACT), was proposed by~\citet{Boggs:2006ACT}. With an energy range of $0.2-10~\unit{MeV}$ and an angular resolution of $1\deg$, ACT should achieve a polarization sensitivity of $1\%-10\%$, depending on the brightness of the source. It could also increase by orders of magnitude the number of Supernovae, AGNs, GRBs, and Novae detected when compared to COMPTEL.

\item{\bf{PANGU}} PAir-productioN Gamma-ray Unit (PANGU) is a proposal for a high angular resolution telescope capable of detecting and measuring polarization for gamma rays in the $\sim10\unit{MeV}-\sim\unit{1GeV}$ energy range~\citep{Wu:2014}. It was suggested as a cooperative mission between the European Space Agency (ESA) and the Chinese Academy of Science (CAS). By using a large number of thin active silicon micro-strip detectors (MSD) tracking layers, PANGU will be able to accurately reconstruct the electron positron pair tracks. The size of the detector is about $80\times80\times 90$ cm, consisting of 100 stacked Si MSDs. In addition, a magnetic spectrometer will be used to determine the energy of the electron and positron. Since the nuclear recoil is expected to be negligible, this translates into a determination of the energy of the incident photon. This novel tracker design is expected to provide unprecedented angular resolutions for sub $\unit{GeV}$ gamma rays, and as such could also serve as a polarimeter.

\item{\bf{POLAR}} One of the most intriguing astrophysical phenomena for which insights into the emission mechanisms could be provided by polarimetry studies are Gamma Ray Bursts. To date, there are only a very limited number of measurements of polarization of the prompt emission of GRBs, and typical errors are large. This is due to the fact that most of those measurements are taken with instruments that were not designed to be used primarily as polarimeters, and hence were not calibrated on the ground. Somewhat surprisingly, there are very few instruments dedicated solely to this aim, even in the somewhat more accessible hard X-ray regime. One such example is POLAR~\citep{POLAR:2016}, a mission that has the aim of performing high precision measurements of the polarization of energetic photons from GRBs in the range of $50~\unit{keV}\sim 500~\unit{keV}$. With a very large FOV, covering about 1/2 of the sky, initial estimates based on ground tests and Monte Carlo simulations indicated that POLAR should be able detect about 50 GRBs per year and to perform high precision measurements of polarization degree for about 10 of those. After being launched on-board the Chinese space laboratory Tiangong-2 in September 2016, POLAR has detected more than 50 GRBs in its first half year of operation in orbit, exceeding initial estimates. About 10 of those GRBs were bright enough to allow detailed polarization studies~\citep{POLAR:2018}. Unfortunately, in April 2017, the instrument stopped taking data due to a problem with the High Voltage power supply, and attempts to recover the system are still ongoing. In spite of this setback, POLAR has demonstrated the usefulness of a wide FOV, high sensitivity gamma ray polarimeter dedicated to the study of GRBs.

\item {\bf{TIGRE}} The Tracking and Imaging Gamma Ray Instrument (TIGRE) Compton telescope was developed by a team at the University of California, Riverside \citep[e.g][]{ONeill:2003TIGRE,Bhattacharya:2004TIGRE} to observe cosmic gamma-rays in the low and medium energy range ($0.1 -100 \unit{MeV}$). This telescope is capable of detecting both Compton interactions and electron-positron pair production. However, the calorimeter design of TIGRE enhances the instrument as a gamma-ray polarimeter only the energies below  $2 \unit{MeV}$.  This balloon-borne telescope for  gamma-ray observations in the MeV energy range uses multi-layers of thin silicon detectors to track and measure the energy losses of Compton recoil electrons. The telescope was tested successfully in 2010 on a 57 hours flight, having being launched on a stratospheric balloon from the Australian Balloon Launching Station, Alice Spring, Australia. At this stage we are unaware of possible future science missions.

\end{itemize}

In the next section, I discuss the main scientific questions that motivated the research and development of so many  instruments sensitive to gamma ray polarization at energies above the pair creation threshold.

\section{Main Motivations for gamma-ray polarimetry in the $\unit{MeV}-\unit{GeV}$ range}\label{Sec:Motivations}

Medium and high energy gamma-ray astronomy has evolved tremendously over the past decade or so. Two pair production space telescopes (AGILE and FERMI/LAT) capable of detecting gamma-rays with energies higher than $\sim100 \unit{MeV}$ being currently operational. Data from those two instruments has greatly improved our understanding of gamma-ray production sites and mechanisms. 

Some of the main discoveries with FERMI-LAT are: detection of certain pulsars that appeared to emit radiation in gamma-rays predominantly, the greatest GRB energy release (GRB 130427), determining the role of supernova remnants as accelerators for cosmic particles, the study of the extragalactic gamma-ray background (EGB), and establishing that unresolved non-blazar active galactic nuclei (AGNs) can only be responsible for about $1/3$ of the entire Cosmic Gamma-ray Background (CGB) flux above $100 \unit{MeV}$~\citep[cf.][]{FERMI:2010EGB,Inoue:2011}, and the Gamma/X-ray bubbles that extend about $8\unit{kpc}$ above and below the Milky Way galactic plane \citep{Su:2010}. The italian telescope AGILE has been used to detect and study numerous blazars (very high energy AGNs), galactic gamma-ray production sites such as the Carina region \citep{Tavani:2009}, the Cygnus region microquasars \citep[eg.][]{Tavani:2009Cygn}, the Vela Pulsar Wind Nebula \citep{Pellizzoni:2009}, terrestrial gamma-ray flashes up to $100 \unit{MeV}$, etc. Even if there are more and more high energy gamma-ray resolved sources detected and catalogued, we still lack a complete understanding of the production mechanisms in some of the most common astrophysical sites for gamma-rays such as AGNs or GRBs. 

It is commonly argued that high energy gamma-ray polarimetry could be used to deepen our understanding of the universe in multiple ways. First, it could be used as a very stringent probe of the various proposed mechanisms for gamma-ray emission in GRBs, AGNs, blazars, pulsars. This is possible since different models will predict a significantly different level of linear polarization, as we shall see in more detail in the following subsections. Moreover, high energy gamma-ray polarimetry has the potential to probe fundamental physics questions, such as the nature of Dark Matter, axions, and violations of Lorentz Invariance in our universe. In Sec.~\ref{SSec:ProdMech}, I reviewed the main production mechanisms for galactic and extragalactic gamma-rays, with a particular emphasis on the various polarization signatures of each. Sections~\ref{Sec:Blazars} to~\ref{Sec:Axion} are dedicated to discussing in more detail how polarization measurements for gamma rays in the $\unit{MeV}-\unit{GeV}$ range can help deepen our understanding of Nature.

\subsection{Blazars}\label{Sec:Blazars}

Most, if not all galaxies, have a central supermassive black hole (SMBH) that can power, via accretion and other associated mechanisms, extremely bright emission in the optical and across a variety of other spectral bands. When this phenomenon happens, the galactic nucleus is called ``active.'' For instance, if an AGN reveals jet-like outflows in the optical band, then it almost always will be a strong emitter of radio waves.

Blazars are a subclass of Active Galactic Nuclei (AGN)  for which the jet emission is almost aligned with the line of sight to us. The emission is typically broad band, with Spectral Energy Distributions (SEDs) probed by the Fermi satellite in the $0.05-50\unit{GeV}$ range~\citep{Ackermann:2015AGN}. Their spectral SEDs usually show two distinct peaks, with the high energy peak invariably found in the gamma-ray regime. The measurement of polarization of X-ray and gamma-ray emission from blazars will help discriminate between various emission mechanisms currently proposed. For instance, the high levels of linear polarization found in the radio and X-ray low energy peak in blazar spectra~\citep[e.g.][]{Hayashida:2012} lead to the commonly accepted scenario responsible for the low energy broadband peak: emission via the synchrotron mechanism from relativistic electrons in the jet in the presence of highly ordered magnetic fields. Depending on the exact location of the low energy, synchrotron ``hump'' in the SED, blazars are further sub divided into Low-Synchrotron Peaked (LSP; $\nu^S_{peak}<10^{14}~\unit{Hz}$) blazars, Intermediate Synchrotron Peaked (ISP; $10^{14}~\unit{Hz}<\nu^S_{peak}<10^{15}~\unit{Hz}$) blazars (which are also sometimes called in the literature Intermediate BL Lac Objects), and High Synchrotron Peaked (HSP $\nu^S_{peak}>10^{15}~\unit{Hz}$) blazaars (or HBL for High-peaked BL Lac objects). As shown in Sec.~\ref{SSec:ProdMech} (see Eq.~\ref{ESRCritFreq}), there is a correlation between the frequency of  the cutoff of the synchrotron radiation and magnetic fields at the location of the emission, or (and) the relativistic boost factor of the emitting electrons.

For the high energy (X and gamma ray) emission from blazars, two fundamentally distinct mechanisms have been proposed: leptonic and hadronic models. In the former case, it is assumed that electrons and positrons (leptons) are responsible, via the IC of the synchrotron radiation, for the high energy spectral peak~\citep[e.g][]{Maraschi:1992}. In the later case (hadronic models), both electrons and protons are accelerated to ultra-relativistic energies, with protons reaching energies above the photon-pion production threshold. The high frequency emission is much more complex in nature now, being composed of synchrotron and Compton radiation from secondary decay products of charged pions, gamma rays from $\pi^0$ decay, and proton synchrotron emission.~\citet{Boettcher:2013} present a review of the main features of both leptonic and hadronic blazar  models. Polarimetry studies in the gamma-ray regime are extremely important, as they can help disambiguate between hadronic vs leptonic models~\citep[e.g.][]{Zhang:2013}. Hadrons from BL Lac blazars jets are one promising candidate for source of the mysterious ultra high energy cosmic rays. As such, a clear identification of the presence of hadrons in the jets of BL Lac blazars help solve this longstanding problem of astrophysics.

In the case of leptonic models the high energy ``hump'' in the SEDs of balzars is thought to originate from Inverse Compton (IC) scattering off of the same electrons responsible for producing the lower energy, synchrotron radiation peak in the broadband spectra. If this scenario is correct, and the seed photons for the Inverse Compton process are the already polarized synchrotron emission lower energy photons, then one would expect a lower but potentially measurable degree of polarization in high energy, gamma-ray peak emission from blazars~\citep{Zhang:2013}.  Conversely, if no detectable polarization is found for the up-scattered photons, this would imply a different origin for the high energy emission than the commonly considered Inverse Compton up-scatter of synchrotron emission. For example, another source of seed photons for the IC scattering could be the accretion disk, via direct or reprocessed emission. This external Compton radiation (EC) is not expected to be polarized since accretion disk emission is unpolarized~\citep{Zhang:2013}. For more details on the expected polarization signatures of inverse Compton emission and implications for Blazar observations, see~\citet{Krawczynski:2012}.

\subsection{Gamma-Ray Bursts}\label{Sec:GRBs}
GRBs are the brightest electromagnetic events known to occur in the universe. However, the exact emission mechanisms are still unknown, as currently there is  no single accepted theory for the powerful ``engines'' behind GRB phenomenon, despite the fact that the first GRB was observed more than 50 years ago by the Vela satellites. The Burst and Transient Source Experiment (BATSE) experiment onboard the Compton Gamma Ray Observatory (CGRO) discovered about two such events per month during its nine year mission. Spectral data from BATSE helped build a unifying picture, with most GRB spectra peaking in the $0.1-1\unit{MeV}$ range, with an enormous energy outflow collimated in a narrow highly relativistic jet. With typical Lorentz factors ($\Gamma$) often in excess of 100, GRBs are the source of the most relativistic jets known. Thus, GRBs are probes of some of the most extreme conditions in the universe. The FERMI/LAT team has recently published a catalogue of LAT-detected GRBs during its first ten years of operations. A total of 186 GRBs are found, with 91 of them showing emission in the $30-100~\unit{MeV}$, and 17 of them being detectable only in this narrow band~\citep{FERMI_LAT:GRB:2019}. This further motivates the need for an instrument which could cover the $~\unit{MeV}$ sensitivity gap, in both polarimetry and spectroscopy.  

Regarding their formation, GRBs have been observed as far back as redshifts of $z\sim 9$, making them powerful probes of the young universe. As I explain in more detail in Sections~\ref{Sec:LIV} and~\ref{Sec:Axion}, the study of polarization of radiation from GRBs is an invaluable tool for probing our understanding of fundamental physics. Since GRB spectra peak at energies typically below $1\unit{MeV}$, Compton based polarimeters (or Compton telescopes sensitive to polarization) have already detected a high degree of polarization of the X-ray emission for a limited sample of very bright GRBs, as discussed in Sec.~\ref{Sec:Instruments}. There is hope that in the near future the number and statistical significance of such detections will be dramatically increased, as dedicated instruments are being built or researched and developed.

GRBs fall under two broad categories, depending on their duration. Short GRBs, lasting less than a couple of seconds, are thought to be the consequence of compact binary mergers, such as NS-NS or NS-BH, whereas collapse of massive stars is thought to be responsible for long GRBs, lasting up to a couple of minutes. After this prompt initial burst, a long lasting, multi wavelength afterglow sets in. Recently the LIGO-VIRGO collaboration discovered gravitational waves from binary neutron star (BNS) mergers~\citep[GW170817][]{Abbot:2017}. A gamma ray counterpart of this event has been observed with Fermi Gamma Ray Burst Monitor (GBM)~\citep[GRB 170817A][]{Goldstein:2017}, thus making the study of short GRBs extremely compelling, as complementary probes of the merging event. Moreover, this milestone result strengthens the hypothesis that short GRBs and the BNS mergers are linked. 

The remainder of this section summarizes how  polarimetry could help address and potentially solve major open questions related to the GRB phenomenon. One candidate for the mechanism behind the prompt GRB emission is synchrotron radiation from particles carried away from the central engine, and in this case one expects a degree of polarization dependent on the ordering of the magnetic fields. For instance, in the case of the ordered field model (SO), a toroidal magnetic field is expected to produce a highly polarized emission. A tell tale signature of this model is the non-uniform degree of polarization, which would be detectable as a time variable polarization angle for the photons emitted in the line of sight~\citep{Lyutikov:2003,Granot:2003}. For random field models, one naturally expects a low degree of measured polarization, if the viewing angle is on-axis, since the polarization angles are axisymmetric along the line of sight. However, if the viewing angle is off-axis, there will be some residual polarization fraction (as large as $50\%$), due to a non-precise cancellation of the polarization vectors~\citep{Granot:2003,Toma:2009}. In contrast to synchrotron class of models, photospheric models assume that the gamma-rays are emitted radiatively from a photosphere and then are beamed towards jet. The linear degree of polarization is expected to be correlated to the luminosity, in this class of models, with a theoretical upper limit of $40\%$, as found by~\citet{Beloborodov:2011}.

Since many of those models can produce similar polarization signatures for individual GRBs, it is impossible to rule them out based on single observations alone. A statistically significant disambiguation between those models could be made via polarimetry in conjunction with spectroscopy~\citep[e.g.][]{Toma:2009}, if a large enough sample of GRBs for which polarization fraction and angle are measured and correlated against other parameters, such as peak energy or duration. GRBs are still poorly understood phenomena, for which a multi-directional approach, combining gravitational wave data (where available) to spectroscopic and polarimetric measurements is particularly useful.

\subsection{Pulsars}\label{Sec:Pulsars}

The first pulsar was discovered in 1957 by Jocelyn Bell, a graduate student at Cambridge University at the time. She found an object that emits in a very narrow band of radio frequencies, at an extremely regular interval. Due to the peculiarities of this signal, it was speculated that it might be of extraterrestrial intelligent origin. However, the discovery of the second such source, in a different galaxy lead to a more rigorous hypothesis, that of a new class of stars, never before observed. We now know that pulsars are  rotating neutron stars which  which emit beams of  particles at opposite poles, and as a consequence their rotation rate decreases over time. The Fermi LAT satellite revolutionized the study of pulsars, by detecting more than 230 of them, classified in several subclasses, as of April 2019~\footnote{For an updated list see \url{https://confluence.slac.stanford.edu/display/GLAMCOG/Public+List+of+LAT-Detected+Gamma-Ray+Pulsars}}. The spectra of most gamma ray pulsars observed presents exponential cutoffs in the $\unit{GeV}$ range, favoring outer gap models~\citep{Cheng:2000}, where the emission originates in the outer magnetosphere. In contrast, the polar cap models~\citep{Ruderman:1975} predict much sharper cut-offs, and are therefore now disfavored. The observed pulsating radiation is only a small part  of the energy lost by pulsars, with most of it being carried away in the form of a magnetized relativistic wind. In some cases, such as for example the Crab nebula, a highly collimated jet and a circumstellar torus are also observed~\cite{Weisskopf:2000}.

Soft gamma ray pulsars are a class of pulsars reaching maximum brightness in the $\unit{MeV}$ band, as opposed to the $\unit{GeV}$ range, characteristic for all other classes of pulsars. Due to their soft spectrum, only a few of them were observed by Fermi LAT. As such, their spectra is still poorly understood. Thus, no compelling emission mechanism has emerged, although there is the suggestion that soft-gamma ray pulsars are just regular pulsars for which the $\unit{GeV}$ part of the beam is missed by our line of sight~\citep{Wang:2012}. Both spectroscopic and polarization data for a large sample of soft gamma rays pulsars will be required in order to fully test this or alternative hypotheses for soft gamma ray pulsar emission.

The discovery of peculiar classes of X-ray pulsars such as the soft gamma repeaters (SGRs) and the anomalous X-ray pulsars (AXPs), lead to the Magnetar hypothesis: a ultra-magnetized ($B\approx 10^{14}$ G) neutron star powered by magnetic energy~\citep[For a recent review, see][]{Turolla:2015}. Their SEDs are measured only up to hard X-rays, with upper limits at higher energies imposed by CGRO Comptel observations. The emission suggested mechanisms involve reprocessing of thermal photons emitted by the star via resonant Compton scattering (RCS). The scatterers are provided by charges moving in a ``twisted'' magnetosphere~\citep{Nobili:2008}. Albeit challenging, in view of the relatively low expected photon fluxes, hard X-ray and soft gamma ray polarimetry for Manetars will provide insights into the geometry of the region where the currents flow and the velocity distribution of the particles responsible for the processing of thermal radiation via the RCS effect.

In general, gamma ray polarimetry studies for pulsars could provide crucial insights into the neutron-star magnetic fields and the region in the magnetosphere where acceleration of particles takes place, thus even further refining the plausibility of various emission mechanisms. It should be noted that in the case of pulsars, since they emit over a broad range of frequencies, polarimetry studies over the entire electromagnetic spectrum would provide unprecedented diagnostic tools.

\subsection{Gamma-ray binaries}\label{Sec:APBinaries}

Gamma ray binaries are systems consisting of a massive star and a compact object, where gamma rays dominate the SED, which peaks at energies in the $1-100\unit{MeV}$ range. For a review of gamma ray binaries and related systems, see~\citet{Dubus:2013}. The nature of their emission is still largely unknown, mostly due to a very limited sample size, and the the very little amount of spectral or polarimetric data, since the peak SEDs overlap with the $\unit{MeV}$ sensitivity gap. As future instruments are predicted to close this gap in the near future, it is expected that in our Galaxy alone we will be able to find anywhere between 50 and 200 such gamma ray binaries~\citep{Dubus:2017}. Most likely, the dominant radiation mechanisms at play for gamma ray binaries are synchrotron emission, responsible for the low (below $\unit{MeV}$) energy part of the SED, and inverse Compton (IC) scattering, of stelar photons, dominant in the very high (above $\unit{GeV}$) regime~\citep{Bosch-Ramon:2009}. It is not yet clearly known which mechanisms dominate in the intermediary $\unit{MeV}-\unit{GeV}$ region, as the broad-band spectrum of gamma-ray binaries can generally be fitted equally well by models where the soft and high energy gamma ray emission is synchrotron or IC dominated. This degeneracy can be broken by measurements of the polarization parameters, in view of the distinct polarimetric signatures of IC and synchrotron radiation, as shown by~\citet{Zdziarksi:2010}. When there is significant emission in the $\unit{TeV}$ range, such as in the case of the gamma ray binary LS I +61 303, then models where synchrotron emission dominates in the intermediary regime are preferred, in view of the sharp high energy cutoff of the inverse Compton component. However, including another spectral component responsible for the $\unit{TeV}$ emission, would bring those two models back in agreement at a spectroscopic level. Therefore polarimetry studies are essential for understanding the exact transition between the low and very high energy regimes of the SEDs.

Once a disambiguation between synchrotron or IC dominated $\unit{MeV}-\unit{GeV}$ emission from gamma ray binaries is made, additional important lessons can be learned. For instance, the synchrotron emission is typically limited to energies below $\sim 100\unit{MeV}$. However, this limit can be exceeded in certain circumstances, such as for instance the Crab Nebula. The determination of a synchrotron component exceeding the $\sim 100\unit{MeV}$ in the context of a gamma ray binary limit would indicate highly relativistic motions, or an additional spectral component contaminating the emission. Other open questions that could be addressed by polarimetric and spectroscopic studies, in the context of gamma ray binaries, include the study of particle acceleration, outflows, and wind launching mechanisms.

\subsection{Supernovae Ia}
It has been recently pointed out in~\citet{Churazov:2018} that gamma ray polarimetry could be useful in the study of SNIa. Specifically, when the scattering of narrow gamma ray lines in the SNIa ejecta is considered, one finds that the degree of polarization and the scattering angle are correlated, as expected. Although the level of asymmetry, and thus the detection prospects, are still an open question, the possibility of using gamma ray polarimetry in the study of SNIa phenomena is a very intriguing avenue.  

%%%%%%DM Indirect Detection. READ AND GHANGE%%%%%%%%%%%%%%%%%

\subsection{Indirect Dark Matter Detection}\label{Sec:DM}

In 2012, analysis of the Fermi data available at the time by~\citet{Bringmann:2012} lead to the identification of a marginally statistically significant monochromatic line, at about $\sim 130\unit{GeV}$, in the gamma ray spectrum from the center of the galaxy. Soon afterwards, the Fermi-LAT team used 3.7 years of data in the $5-300\unit{GeV}$ range to search for monochromatic gamma-ray lines in a set of five circular regions of interest (ROI) centered on the Galactic center picked in such a way to maximize sensitivity to various theoretically motivated DM density profiles~\citep{Ackermann:2013GRLine}. The authors do not find any globally significant lines; however, multiple locally significant signals are found. At $133\unit{GeV}$, a line-like signal with a local significance of $3.3~\sigma$ is present originating from the inner most region of interest. Yet, the global significance decreases to $1.5~\sigma$ when one takes into account all possible locally significant signals in all five ROIs. This is the same feature reported previously in the unprocessed PASS 77 data at $130\unit{GeV}$ by various groups \citep{Bringmann:2012,Weniger:2012}. The shift is attributed to the improved calibration. Since there were no globally significant lines found, the authors use their ``null detection'' results to place flux upper limits on monochromatic sources, which are then translated to upper limits on DM self annihilation cross section or DM decay lifetime lower limits. Results of the extended search for gamma-ray lines of lower energies, from $100\unit{MeV}$ to $10\unit{GeV}$, were presented by the Fermi team in ~\citet{Ackermann:2013GRLine}. Annihilation features at those lower energies are expected in decaying Gravitino DM models or from DM annihilation to two photons~\citep{Albert:2014}. At energies below $\sim15\unit{GeV}$, systematic errors are dominant for the Fermi-LAT telescope. A careful treatment of systematics in their analysis leads to the result that no globally significant lines are detected, attributing the features found at $133\unit{GeV}$ or $327\unit{MeV}$ to a combination between systematics and statistical fluctuations of the backgrounds.

It is worth mentioning that all of those results are dependent on the various backgrounds subtraction. Therefore, a better understanding of the complicated gamma-ray backgrounds from the center of the galaxy is necessary before the question of high energy gamma-ray lines as potential signals of DM annihilations in our galaxy could be settled. The most recent data from both the H.E.S.S.~\citep{Abdallah:2018} and the Fermi-LAT~\citep{Ackermann:2015} telescopes do not find any significant gamma ray emission lines. As such, those null results place tight constraints on the annihilation cross sections for mono-energetic gamma ray lines.  Albeit there is still hope that, as more and more data is collected, a positive identification of a sharp spectral line could be made. This would constitute a ``smoking gun'' for DM annihilations directly to photons, but this possibility seems more and more remote in view of recent data. However, not all hope of indirect DM detection via gamma ray signals is lost, as the search for a more broadband excess has yielded very promising results, as we shall shortly see.  In this case polarimetry could play a crucial role in our attempt to disambiguate between DM annihilation signals the complicated astrophysical backgrounds from very energetic inner regions of our Galaxy.

Diffuse Galactic gamma-ray emission (DGE) is a form of gamma-rays generated in the ISM by interactions of high energy cosmic rays and is the dominant form of radiation observed at energies higher than $\sim100\unit{MeV}$. Bremsstrahlung, inverse-Compton scattering of background photons, and pion production in proton-nucleon scattering events all lead to the diffuse continuum emission. Detailed observations of the DGE have been published by the Fermi-LAT team~\citep{Ackermann:2012DGE, Fronasa:2016} and allow for a thorough study of cosmic rays in the interstellar medium. At energies above a few GeV, most diffuse emission models based only on non-thermal production and standard particle reactions sources of gamma-rays {\it{underpredict}} the total Fermi-LAT measured fluxes originating from the inner $150 \unit{pc}$ of the Galactic Center. This hints towards a possible exotic particle source of gamma-rays of non-astrophysical origin. Primary candidates are DM annihilations or decay in the dense DM environment at the galactic center.  The technique of indirect detection of DM based on decay or annihilation broad spectral signals has been explored in the literature in numerous articles. For a recent review, see~\citet{Gaskins:2016}. Research in this direction was boosted in 2008 when PAMELA confirmed a positron excess at energies above $\sim 10\unit{GeV}$, when compared to the standard cosmic ray model \citep{Adriani:2008Pamela}. This result was later supported in 2011 when the FERMI Gamma-Ray Space Telescope confirmed the previously found positron excess \citep{FermiLAT:2011Positrons}. More recently the AMS-02, an experiment with exquisite statistics, has mapped the positron fraction excess to energies up $\sim 300\unit{GeV}$ confirming previous indications that there is an excess of positrons compared to standard astrophysical backgrounds in our galaxy \citep{Aguilar:2013:AMSPositron}. The last piece of data that firmly established this as a real, physical signal, beyond the shadow of a doubt, is the measurement published by the PAMELA collaboration in \citet{Adriani:2013Pamela}. One possible source for such a large positron excess in the cosmic rays could be dark matter particles with masses $\sim \unit{TeV}$ and greatly boosted annihilation cross sections compared to the Weakly Interacting Massive Particle (WIMP) standard thermal relic cross section. This boost is necessary in order to explain the large signals observed, compared to known backgrounds. Even if not excluded, those models are in tension with other hints of detection of dark matter that favor a much lighter mass for the WIMP, and a cross section closer to the thermal relic value of $\langle\sigma v\rangle\sim 10^{-26}\unit{cm}^3\unit{s}^{-1}$. Either Dark Matter is not made of one unique particle, or, more naturally, some of those signals are actually of astrophysical origin. For example, cosmic rays might be generating positrons via interactions with ambient matter in supernova shock waves, a proposal put forth by \citet{Blasi:2009} and \citet{Mertsch:2011}, extending the standard supernova remnant (SNR) model for the origin of galactic cosmic rays (GCRs). Another leading astrophysical source of positrons could be one or more mature and energetic relatively close pulsars \citep{Linden:2013Pulsars}. If such pulsars are identified by searching for anisotropies in the positron signal, the corresponding gamma-ray spectrum and its polarization could be used to strengthen this hypothesis.

Returning to the subject of the diffuse galactic emission (DGE), we note that multiple groups, using the publicly available Fermi-LAT data, have independently claimed detections of an extended excess in the flux staring at a few GeV and coming from the inner few degrees of the GC region~\citep[e.g.][]{Goodenough:2009,Hooper:2010,Boyarsky:2010,Hooper:2011,Abazajian:2012,Gordon:2013,Abazajian:2014}. All the previously mentioned studies link this signal to Dark Matter annihilations, not ruling out millisecond pulsars as a possible alternative astrophysical explanation \citep{Abazajian:2010,Hooper:2010,Hooper:2011,Abazajian:2012,Gordon:2013,Abazajian:2014,Fermi-LAT:2017MSP} or cosmic rays interactions with baryonic gas \citep{Hooper:2010,Hooper:2011,Abazajian:2012,Gordon:2013}. In 2012 the Fermi-LAT team concluded, through their own analysis \citep{Ackermann:2012DM}, that the excess signal observed in the DGE coming from the Galactic Center (GC) does not require the addition of a Dark Matter annihilation or decay component to the flux. Their result is, however, based on a model of the high-energy galactic diffuse emission that is limited by the presence of various, unaccounted for, residuals, at the $\sim30\%$ level! Adopting a conservative attitude, the Fermi-LAT team used their data to place constraints on Dark Matter models. This problem was revisited by \citet{Daylan:2014}, who generated high resolution gamma-ray maps by applying cuts to the Fermi event parameter CTBCORE. Those new and improved maps allow for a more robust separation of backgrounds, and indicate the presence of gamma-ray excess at a few $\unit{GeV}$, which is highly statistically significant. The flux can be very well fitted by a $30-40\unit{GeV}$ WIMP annihilating mainly to $b\bar{b}$ with a cross section $\langle\sigma v\rangle=(1.4-2)\times 10^{-26}\unit{cm}^3\unit{s}^{-1}$. This signal is distributed with an approximate spherical symmetry around the GC and extends out to angles of at least $10\deg$. In light of the extended nature of this signal, the previously commonly accepted $\sim 1000$ population of millisecond pulsars as a possible alternative explanation becomes disfavored. Even prior to this paper, it was shown by \citet{Hooper:2013} that no more than $\sim5-10\%$ of the anomalous DGE from the Inner Galaxy can be attributed to pulsars, as Fermi-LAT should have already resolved a much greater number of such objects. The other possible explanation -cosmic rays interacting with gas in the Inner Galaxy- is also disfavored considering the morphology of the signal \citep[e.g.][]{Linden:2012,Macias:2013}. This anomalous excess with a flux peaking at a few GeV consists of $\sim 10^4$ gamma-rays per square meter per year above $1 \unit{GeV}$.

In addition to an excess positron signal, several groups recently identified an anomalous excess in anti-proton signal reported by the AMS-02 experiment~\citep{AMS02:2016}. \citet{Chlois:2019} shown that this excess spectrum could be due to a $\sim 70\unit{GeV}$ DM particle annihilating to $b\bar{b}$ with a cross section of $\sigma v\approx2\times10^{-26}\unit{cm}^2/\unit{s}$. Most intriguingly, the same range of DM models could are favored to explain the gamma ray GeV excess from the center of the Galaxy observed in the FERMI-LAT spectrum, thus strengthening the case of a DM interpretation of all of those anomalous signals. As of today, however, there is no consensus yet on this issue. As already pointed out, those excess signals could be of either astrophysical origin (i.e. pulsars), denoting an incomplete knowledge of the relevant backgrounds, or they could be indirect signals of Dark Matter. Polarimetry could definitely help discriminate between those two scenarios. The astrophysical background gamma rays, even if enhanced by a MSP population to account for the DGE excess at a few $\sim\unit{GeV}$, is not expected to show any significant degree of polarization. On the other hand, certain class of DM models predict a significant degree of linear~\citep{Huang:2018} or circular~\citep{Ibarra:2016,Kumar:2016,Bonivento:2017,Boehm:2017,Elagin:2017} polarization of the photon flux from DM annihilations. The measurement of this polarization is, however, very challenging, as it involves separating a relatively small signal from a large, already polarized, background.

Another strategy, proposed by \citet{Baltz:2006}, would be to search for similar signals from dwarf spheroidal satellite galaxies, where backgrounds are much smaller. The Fermi-LAT team, by using 6 years of their data, find that none of the 25 targeted Milky Way dwarf spheroidal galaxies present significant fluxes to claim detection in gamma-ray in the $500\unit{MeV}-500\unit{GeV}$ regime~\citep{Ackermann:2015DS}. Those null results are used to place strong constraints on DM annihilation cross sections, that lie below the canonical thermal relic value.

\subsection{High Energy Gamma-Ray Polarimetry as a probe Lorentz Invariance}\label{Sec:LIV}

 One of the possible effects induced by quantum gravity would be the presence of small, but potentially detectable, Lorentz or CPT violating terms in the effective field theory action. Those terms lead to a macroscopic bi-refringence effect of the vacuum \citep[i.e.][]{Jacobson:2005}. In an effect similar to the Faraday rotation, but of a radically different nature, the linear polarization direction would be rotated for monochromatic gamma-rays. The rotation angle can be expressed as:
\begin{equation}
\theta\simeq\frac{\xi E^2t}{2M_P}.
\end{equation}
Here $E$ is the photon energy, $t$ is the propagation time, $M_P$ is the Planck mass and $\xi$ is the dimensionless parameter that characterizes the strength of the Lorentz Violating terms in the effective field theory action. This effect has a strong energy dependence and is larger for sources that are farther away, since it is linear in the propagation time, $t$. For gamma-rays with a broad spectral feature, i.e. not monochromatic lines, the effect of the vacuum birefringence wold be to dilute the polarization due to the energy dependence of the rotation angle. Up to now, no such vacuum bi-refringence has been observed. Therefore, one can only place upper bounds for the value of $|\xi|$. The present limit of $|\xi|\lesssim 3.4\times 10^{-16}$, comes from the polarization measurement of the GRB 061122~\citep{Gotz:2013GRB}. Extending polarization sensitivity to higher energies could lead to a detection of the vacuum birefringence, which would have extraordinary implications on fundamental physics. It would be the first ever recorded signature of Quantum Gravity! Even in the case of null detection, one could significantly improve the present limits on the LIV parameter $\xi$, therefore placing constraints on theories of Quantum Gravity that induce Lorentz Violating terms.

\subsection{High Energy Gamma-Ray Polarimetry as a probe for axions}\label{Sec:Axion}

GRB polarimetry could be used to search for the axion, a hypothetical particle introduced to solve the strong CP problem of QCD. In technical terms, the axion is the would-be Nambu Goldstone boson of chirial $U(1)$ symmetry introduced by~\citet{Peccei:1977} in order to explain the smallness of the CP violating $\Theta$ term in QCD. This particle, with an very small predicted mass, of the order of $10^{-6}-1 \unit{eV}$ has not yet been detected, leaving the strong CP problem as one of the most important theoretical problems of the Standard Model for which no definitive solution has been found yet. In general, axion like particles (ALPs) are predicted by many extensions of the Standard Model. They are very light, neutral pseudo-scalar bosons, that couple to two photons, in the same manner as the Peccei Quinn Axion would. The interaction term can be written as:

\begin{equation}\label{Eq:Axion}
\Lagr_{a\gamma}=-\frac{1}{4}g_{a\gamma\gamma}F_{\mu\nu}\widetilde{F}^{\mu\nu}a,
\end{equation}

where a is the axion (or ALP) field, $g_{a\gamma\gamma}$ is the axion-photon coupling strength and $F_{\mu\nu}$ is the usual electromagnetic tensor. This coupling is one of the most important phenomenological property of axions, as it allows axions to be converted to photons in the presence of strong electro-magnetic fields \citep{Dicus:1978,Sikivie:1983}. Based on null detection of photon-axion conversion in the ionized core of the Sun the CAST experiment at CERN placed an upper bound of $g_{a\gamma\gamma}<0.66\times10^{-10}\unit{GeV}^{-1}$ for axions lighter than $0.02\unit{eV}$.

This mixing term of Eq.~\ref{Eq:Axion} also acts as a polarizer, leading to vacuum birefringence (change of linear into elliptical polarization) and dichroism (a rotation of the polarization). Consider a beam of initially linearly polarized photons of frequency $\omega$ and wave-vector $\mathbf{k}$ propagating in a uniform magnetic field $\mathbf{B}$. Due to birefringence, the polarization orientation is rotated by an angle that can be approximated by the following formula~\citep{Maiani:1986}:

\be
\epsilon=N\left(\frac{g_{a\gamma\gamma}B\omega\sin\left(\frac{m_a^2L}{4\omega}\right)}{m_a}\right)^2\sin2\phi,
\ee

where $N$ is the number of passes through the magnetized region of length $L$, and $\phi$ is the angle between $\mathbf{k}$ and $\mathbf{B}$. The frequency dependence of the rotation angle leads to a misalignment of polarization planes of low and high energy events. As such, to the the dilution of the measured polarization fraction for astrophysical sources in which the emission region is strongly magnetized, such as GRBs.  Those kind of effects are used by \citet{Rubbia:2008} to places the following constraint on the axion-photon coupling: $g_{a\gamma\gamma}\lesssim 2.2\times 10^{-11}\unit{GeV}^{-1}$ for an axion mass of $10^{-3}\unit{eV}$. This bound scales like $1/\sqrt{E}$, therefore polarimetry of GRBs at higher energies would lead to even tighter constraints.

It is amazing how the measurement of an astrophysical phenomena, such as polarization of gamma-rays can be used to probe fundamental physics such as Lorentz Invariance or properties of the axion (or ALPs).

\section{Summary and Outlook}\label{Sec:Outlook}

During the past few decades, a new era in high energy astrophysics has been ushered in by major experimental advances. New, all sky surveys and high sensitivity probes in the X-ray (e.g. Neil Gehrels Swift Observatory), as well as high and very high energy gamma-rays (e.g. Fermi-LAT and H.E.S.S) have all contributed to broadening our understanding of the universe. The picture that emerges is one where non-thermal emissions are ubiquitous, powering sources such as Blazars, GRBs, pulsars, gamma-ray binaries, etc. Although theoretical models for explaining each of those phenomena exist, they are all far from being fully understood, as the data available can be explained by competing models that are based on very different assumptions. For instance, in the case of Blazars both leptonic and hadronic models are viable scenarios; for GRBs, the prompt emission can be explained by synchrotron radiation or by photospheric models. Additionally, the degree of order of the magnetic fields thought to be responsible for the synchrotron emission is poorly constrained. In the case of gamma-ray binaries, the transition between the synchrotron and the Inverse Compton parts of the spectrum is poorly constrained. For each of those phenomena, there is virtually no gamma-ray data in the $\unit{MeV}-\unit{GeV}$ sensitivity gap, which is one of the reasons for the aforementioned incomplete understanding of their nature. In the near future the gamma-ray $\unit{MeV}-\unit{GeV}$ sensitivity gap should be bridged by instruments that take advantages of advances in detector technologies such as Silicon strip detectors, Cesium Iodine scintillators, and Cadmium Zinc Telluride detectors. In conjunction with spectroscopy and photometry, gamma-ray polarimetry will  play an important role in disambiguating the aforementioned competing models, since they all are all designed to match available spectroscopic data. As discussed in Sec.~\ref{SSec:Proposals}, there are a significant number of proposals and instruments being built that, in the near future, should be capable of measuring, the linear degree of polarization for photons of astrophysical origin with energies above the pair creation threshold for the first time. This would indeed open a new window into our understanding of the non-thermal emissions in the Universe; in addition, high-energy photon polarimetry could also be used to test our understanding of the nature of Dark Matter (see Sec.~\ref{Sec:DM}), axions (see Sec.~\ref{Sec:Axion}), or as a probe for deviations from Lorentz invariance (see Sec.~\ref{Sec:LIV}), one of the most fundamental symmetries commonly assumed to be realized in nature. Gamma ray polarimetry, in conjunction with photometry and spectroscopy in the $~\unit{MeV}-~\unit{GeV}$ regime, could thus become an invaluable tool for deepening understanding of the $\unit{MeV}$ domain, which remains one of the most under-explored windows on the non-thermal Universe.

\acknowledgements

I would like to thank Branislav Vlahovic for introducing me to the world of gamma-ray polarimetry, and for encouraging me to write this review. I also thank the anonymous referee and Denis Bernard  for their very helpful suggestions and comments. In addition, I would like to thank Eugene Churazov, and Floyd Stecker for pointing out several very relevant papers. And last, but not least, I thank Danielle Lupton, for her many suggestions after proofreading the manuscript. 

\bibliography{RefsPolarization}

\end{document}